\documentclass[11pt]{article}
\usepackage{amssymb}

\newcommand{\be}{\begin{equation}}
\newcommand{\ee}{\end{equation}}
\newcommand{\bd}{\begin{displaymath}}
\newcommand{\ed}{\end{displaymath}}
\newcommand{\bi}{\begin{itemize}}
\newcommand{\ei}{\end{itemize}}
\newcommand{\C}{\mathbb{C}}            
\newcommand{\R}{\mathbb{R}}            
\newcommand{\N}{\mathbb{N}}            
\newcommand{\Dsum}{\bigoplus}          
\newcommand{\dsum}{\oplus}             
\newcommand{\Tens}{\bigotimes}         
\newcommand{\tens}{\otimes}            
\newcommand{\lsymm}{\ell_\mathrm{\small symm}^2}  

\newcommand{\symm}{\mathrm{S}}            

\newcommand{\G}{\Gamma}                
\newcommand{\K}{{\mathcal K}}        
\newcommand{\A}{{\mathcal A}}  
\newcommand{\D}{{\mathcal D}_s}        

\newcommand{\br}{\left\langle}
\newcommand{\ke}{\right\rangle}

\newcommand{\ra}{\rightarrow}

\newcommand{\om}{\omega}

\newcommand{\definitie}{\noindent\textbf{Definition.} }

\newtheorem{lemma}{Lemma}[section]
\newtheorem{theorem}{Theorem}[section]

\newtheorem{proposition}{Proposition}[section]

\def\square{{\vcenter{\vbox{\hrule height.4pt \hbox{\vrule width.4pt 
height1.45ex \kern1.45ex \vrule width.4pt}
\hrule height.4pt}}}}

\def\qed{\hfill$\square$}

\newcommand{\barint}{\hbox{$\int$\kern-0.75\intwidth\vrule width 0.5\intwidth height 2.4pt depth -2pt\kern0.25\intwidth}}
\newlength\intwidth
\setbox0=\hbox{$\int$}
\intwidth=\wd0

\begin{document}
\title{ Symmetric Hilbert spaces \\
arising from species of structures}
\author{\normalsize M\u ad\u alin Gu\c t\u a 
\footnote{E-mail: guta@sci.kun.nl}\and \normalsize Hans Maassen
\footnote{E-mail: maassen@sci.kun.nl}}
\date{\today}
\maketitle

\begin{center}
{\rm Mathematisch Instituut}\\
{\rm Katholieke Universiteit Nijmegen} \\ 
{\rm Toernooiveld 1, 6526 ED Nijmegen}\\
{\rm The Netherlands}\\
{\rm fax}:+31 24 3652140
\end{center}

\begin{abstract}\noindent
Symmetric Hilbert spaces such as the bosonic and the fermionic Fock
spaces over some `one particle space' $\K$
are formed by certain symmetrization procedures performed on the full Fock space.
We investigate alternative ways of symmetrization by building              
on Joyal's notion of a combinatorial species.
Any such species $F$ gives rise to an endofunctor $\G_F$
of the category of Hilbert spaces with contractions
mapping a Hilbert space $\K$ to a symmetric Hilbert space $\G_F(\K)$
with the same symmetry as the species $F$.
A general framework for annihilation and creation operators
on these spaces is developed,
and compared to the generalised Brownian motions of R. Speicher
and M. Bo\.zejko.
As a corollary we find that the commutation relation
$a_ia_j^*-a_j^*a_i=f(N)\delta_{ij}$ with $Na_i^*-a_i^*N=a_i^*$
admits a realization on a symmetric Hilbert space whenever
$f$ has a power series with infinite radius of convergence
and positive coefficients.
\end{abstract}

\section{Introduction}
\label{Secintro}\noindent
Symmetric Hilbert spaces play a role in physics as the state spaces
of many particle systems.
The type of particle dictates the type of symmetrization:
bosons require complete symmetrization
and fermions complete antisymmetrization.

\noindent
More general ways of of symmetrization,
although apparently not realized in nature,
have been studied for their own sake: 
parastatistics \cite{Ohn.Kam.} and
interpolations by a parameter $q\in[-1,1]$
between the above two cases
\cite{Grb., Fiv., Zag., Boz.Sp.2, Maa.vLee.}.

\noindent
All these constructions lead to quantum fields or generalized Brownian motions,
each with their own generalized Gauss distributions
\cite{Boz.Ku.Spe., Boz.Sp.1, Boz.Sp.2, Boz.Sp.3, Maa.vLee.}.
One particularly important case is $q=0$:
the free Brownian motion, exhibiting the Wigner distribution.
This case is related to free independence in the same way as
the case $q=1$ of complete symmetrization is related to ordinary
commutative independence.

\noindent
Although there are results \cite{Spe.2} indicating that these two
are the only notions of independence,
more relaxed conditions such as the weak factorization property \cite{Kumm.},
or pyramidal independence \cite{Boz.Sp.1} are satisfied in a variety
of examples.

\noindent
In this paper we study combinatorial ways of symmetrization.
Our starting point is the following observation. 
The category $E$ of finite sets has as its isomorphy classes the natural
numbers $\N$,
and for each object $U$ in class $n\in\N$ there are $n!$ symmetries.
This leads to the Fock space
\bd
l^2\left( \N,{1\over{n!}}\right)=:\G_E(\C).
\ed
Taking for an annihilation operator $a$ the left shift on this space,
we find that the field operator $X:=a+a^*$ has distribution given by
(\cite{Simon})
\bd
\br\delta_\emptyset,X^n\delta_\emptyset\ke
={1\over\sqrt{2\pi}}\int_{-\infty}^\infty x^n e^{-{1\over2}x^2}\;dx.
\ed
On the other hand,
if we consider the category $L$ of finite sequences
(or linear orderings on a set),
we obtain
\bd
\G_L(\C)=l^2(\N)
\ed
and, since $a^*$ is now the right shift \cite{Voi.Dy.Ni.},
\bd
\br\delta_\emptyset,X^n\delta_\emptyset\ke
={1\over\pi}\int_{-2}^2 x^n\sqrt{4-x^2}\,dx\;.
\ed
We conclude that the Gauss and Wigner distributions are produced by
the concepts of `set' and `sequence'.
Our program in this paper is to generalize the Fock space construction
to such combinatorial concepts as `tree', `graph' and `cycle'.

\noindent
The proper framework for this undertaking turns out to be Joyal's
notion of a {\it combinatorial species of structures} \cite{Joyal}.
These are defined as functors from the category of finite sets with bijections
to the category of finite sets with maps.
Combinatorial species of structures can be viewed as coefficients of the
Taylor expansion of analytic functors \cite{Joyal.2} and lead to Joyal's
notion of a tensorial species,
very close to the our $\G_F(\K)$.
This circle of ideas is introduced in Sections 2 and 3.

\noindent
A natural way to introduce an
annihilation operator into this context is via the operation
of removal of a point from a structure,
which is called {\it differentiation} $F\mapsto F'$ by Joyal.
We thus arrive at operators
\bd
a(k):\G_F(\K)\to\G_{F'}(\K),\quad a^*(k):\G_{F'}(\K)\to\G_{F}(\K),\quad(k\in\K)
\ed
However, these operators can only be added, in order to yield field
operators, if the species $F$ and $F'$ are the same.
This holds in two cases: the species $E$ of sets and the species $E_\pm$
of oriented sets,
related to the Bose and Fermi symmetries.
Natural as this may be,
we cannot move any further if we do not modify the operation of removal of
a point in some way.

\noindent
Now in fact, already in the case of sequences it is required that
{\it the last} point is removed.
In the same way, we may require in the case of trees that only leaves
may be picked off (points that leave the tree connected when removed).
In the case of cycles we may require that the chain, coming from a broken,
cycle must be connected up again.
All of this leads to the study of suitable transformations between
$F$ and $F'$, which are the subject of Section 4.

\noindent
Our approach to symmetric Hilbert spaces and field operators
provides a tool for creating new examples, and is particularly
transparent due to the use of combinatorial objects which are easy
to visualize.
The two examples of ``q-deformations'' appearing in \cite{Boz.Sp.1}
and \cite{Boz.Sp.2} are cast in the form of
combinatorial Fock spaces for the species of ballots and the species of
simple directed graphs respectively.
In  section \ref{Secoperations} we point out the connection between
this combinatorial approach and the one based on positive definite functions
on pair partitions \cite{Boz.Sp.1}.
We describe how the operations between species can be extended to
the weights, illustrating this by examples.

\section{Species of Structures}
\label{Secspecies}
This section is a brief introduction to the combinatorial theory of species of
structures \cite{Ber.Lab.Ler.}, \cite{Joyal},
insofar as is needed here.

\noindent
We are concerned with the different kinds --- or `species' ---
of structures that can be imposed on a set $U$.
The basic idea is that such a species is characterized by the way
it transforms under permutations of the set $U$.

\noindent
It will be convenient to consequently adhere to von Neumann's construction
of the natural numbers according to which $0=\emptyset$ and $n+1=n\cup\{n\}$,
so that the number $n$ coincides with the set $\{0,1,2,\ldots,n-1\}$.

\definitie\cite{Ber.Lab.Ler.}
A \textit{species of structures} is a rule $F$ which\\
(i) produces for each finite set $U$ a finite set $F[U]$,\\
(ii) produces for each bijection $\sigma ~:U\rightarrow V$ a function
$F[\sigma]:F[U]\rightarrow F[V]$.\\
\indent 
The function $F[\sigma]$ should have the following functorial properties:\\
(a) for all bijections $\sigma ~:U\rightarrow V$, $\tau~:V\rightarrow W$,
we have $F[\tau\circ\sigma]=F[\tau]\circ F[\sigma]$, \\
(b) for the identity map $\mathrm{Id}_U :U\rightarrow U$, $F[\mathrm{Id}_U]= 
\mathrm{Id}_{F[U]}$.\\
\smallskip\noindent
The elements of $F[U]$ are called {\it $F$-structures}
on $U$ and the function $F[\sigma]$ describes the
transport of $F$-structures along $\sigma$.
Note that $F[\sigma]$ is a bijection by the functorial property of $F$.

\noindent
We denote by $H_{s}$ the stabilizer $\{\sigma\in \symm (U): F[\sigma](s)=s\}$
of the structure $s\in F[U]$.

\smallskip\noindent
{\bf Examples.}

\noindent
1.
The species $E$ of sets is given by
\begin{eqnarray}
E[U]&=&\{U\}.\nonumber\\
E[\sigma]&:&U\mapsto V\quad\hbox{if}\quad\sigma:U\to V.\nonumber
\end{eqnarray}
Thus the only $E$-structure over $U$ is the set $U$ itself.
The stabilizer of this structure coincides with the whole permutation group
$H_{s}=\symm (U)$.\\

\noindent 2.
The species $L$ of linear orderings: 
\bd
L[U]= \{ f:|U|\rightarrow U\,;\, f~\textrm{bijective} \} 
\ed
where $|U|=\{0,1,2,\ldots,|U|-1\}$ is the cardinality of $U$.
The transport along the bijection $\sigma:U\rightarrow V$ is given by
\bd
L[\sigma](f)=\sigma \circ f.
\ed
The stabilizer of each linear ordering is trivial.
The cardinality of the set of structures $L[U]$ is equal to that
of the permutation group $\symm (U)$. \\

\noindent 3.
The species $\mathcal{C}$ of cyclic permutations:
\begin{eqnarray}
\mathcal{C}[U]&=&\{\pi\in\symm_{U}~|~
\pi^{k}(u)\neq u ~\textrm{for all } u\in U, k<|U|\};\nonumber\\
\mathcal{C}[\sigma]:\pi&\mapsto&\sigma\circ\pi\circ\sigma^{-1}\;.\nonumber
\end{eqnarray}
Each structure $\pi\in \mathcal{C}[U]$ has a nontrivial stabilizer
$H_{\pi}=\{\pi^{k}~|~k<|U|\}$,
the number of structures is
\bd
|\mathcal{C}[U]|=\frac{|U|!}{|H_{\sigma}|}=(|U|-1)!
\ed

\definitie
A species of structures $F$ is called \textit{molecular} if 
the permutation group acts transitively on its structures.
\noindent
A molecular species can be characterized by the conjugacy class
of the stabilizer of any of its structures.
Indeed for $s,t\in F[U]$ and $s=F[\sigma](t)$ we have
$H_{s}=\sigma\circ H_{t}\circ\sigma^{-1}$.
By a well-known combinatorial lemma we have for each structure $s$:
\bd 
|F[U]|\cdot|H_{s}|=|U|! \quad\mathrm{for }\quad s\in F[U]
\ed
\noindent 
In general a species of structure may not be molecular, in which case
it is a sum of species:\\
\smallskip\definitie
Let $F,~G$ be species of structures.
Then their \textit{sum} $F+G$ is the species defined by the disjoint union
\bd
(F+G)[U]=F[U]\cup G[U],
\ed
and the transport along the bijection $\sigma :U\rightarrow V$ is given by:
\begin{displaymath}
(F+G)[\sigma](s)=\left\{\begin{array}{ll}
F[\sigma](s) & \textrm{if $s\in F[U]$} \\
G[\sigma](s) & \textrm{if $s\in G[U]$}
\end{array}\right.
\end{displaymath}
The \textit{canonical decomposition} of a species $F$ is its decomposition
as a sum $F=F_0+F_1+F_2+\cdots$ where $F_n$ denotes the $n-th$ \textit{level}
of $F$:
\begin{displaymath}
F_{n}[U]=\left\{\begin{array}{ll}
F[U] & \textrm{if $|U|=n$} \\
\emptyset & \textrm{if $|U|\neq n$}
\end{array}\right.
\end{displaymath}
The simplest species having 
structures at only one level is the species of singletons $X$:
\begin{displaymath}
X[U]=\left\{\begin{array}{ll}
\{ U\} & \textrm{if $|U|=1$} \\
\emptyset & \textrm{otherwise}
\end{array}\right.
\end{displaymath}
Besides addition,
there is a number of other operations between species
by which to construct new species out of simpler ones. 
Following a standard notation \cite{Ber.Lab.Ler.},
 we use the sum symbol to denote disjoint reunion. \\
\smallskip
\definitie
Let $F,~G$ be two species of structures.
We define the \textit{product} species $F\cdot G$ as:
\bd
(F\cdot G)[U]=\sum_{(U_1, U_2)}F[U_1]\times G[U_2]
\ed
where the sum runs over all partitions of the set $U$ into disjoint parts
$U_1$ and $U_2$.
The transport along the bijection $\sigma:U\rightarrow V$ of the structure
$s=(f, g)\in (F\cdot G)[U]$ is:
\bd
(F\cdot G)[\sigma](s)=(F[\sigma_1](f),G[\sigma_2](g))
\ed
where $f\in F[U_1]$ , $g\in G[U_2]$ and $\sigma_1$, $\sigma_2$
are the restrictions of $\sigma$ to the sets $U_1$ and $U_2$
respectively.

\noindent
The stabilizer of $s=(f,g)$ is
$H_{(f,g)}=H_f\cdot H_g \subset 
\symm(U_1)\cdot\symm(U_2)\subset\symm(U_1+U_2)$.

\noindent
As an example let us consider the n-th power of the species $X$ of singletons:
\begin{displaymath}
X^{n}[U]= \left\{\begin{array}{ll}
\{(u_1,..,u_n)|u_i\in U,~u_i\neq u_j ~\mathrm{for}~i\neq j\} & \textrm{if $|U|=n$} \\
\emptyset & \textrm{otherwise}
\end{array}\right.
\end{displaymath}
It is clear that the species $X^{n}$ and $L_{n}$ are essentially the same.
Indeed there exists a natural bijection between $X^{n}[U]$ and $L_n[U]$:
\bd
(u_1,..,u_n)\mapsto ~\bigl(u:n\rightarrow U:i\mapsto u_i\bigr)\;.
\ed
\textbf{Remark:}
In the language of category theory,
a species of structures $F$ is a \textit{functor} from the category
$\mathbb{B}$ of finite sets with bijections to the category $\mathbb{E}$
of finite sets with functions. \\

\definitie
A \textit{morphism} from the species of structures $F$ to the species $G$
is a natural transformation of functors,
that is a family of functions $m_U:F[U]\rightarrow G[U]$ such that:
\bd
G[\sigma]\circ m_U=m_V\circ F[\sigma]\quad\textrm{for all}\quad\sigma:
                   U\rightarrow V.  
\ed
An \textit{isomorphism} is an invertible morphism.\\

\definitie
The \textit{cartesian product} $F\times G$ of two species of structures
$F$ and $G$ is given by:
\begin{eqnarray}
 (F\times G)[U] & = & F[U]\times G[U]\\
(F\times G)[\sigma](f,g) & = & (F[\sigma](f), G[\sigma](g))\nonumber
\end{eqnarray}
\noindent
The canonical decomposition of the cartesian product is:
\be
F\times G =\sum_{n=0}^\infty F_n \times G_n.
\ee
A structure $(f,g)\in (F\times G)[U]$ has the stabilizer
$H_{(f,g)}=H_f\cap H_g\subset\symm (U)$.\\

\noindent
An operation which will play an important role later is the derivation.\\

\definitie
The \textit{derivative} $F'$ of a species $F$ is a species whose set of
structures over a finite set $U$ is given by:
\bd
F'[U]=F[U\cup\{U\}]
\ed
and  $F'[\sigma](s)=F[\sigma^{+}](s)$ where
$\sigma^{+}:U\cup\{U\}\rightarrow V\cup\{V\}$ is the extension of
$\sigma:U\rightarrow V$:
\begin{displaymath}
\sigma^{+}(x)=\left\{\begin{array}{ll}
\sigma(x) & \textrm{if $x\in U$,} \\
V & \textrm{if $x=U$.}
\end{array}\right.
\end{displaymath}

\noindent\textbf{Remark:}
The term $\{U\}$ in $U\cup\{U\}$ is just any additional point,
not belonging to $U$.
In particular for $U=n$ we have $U\cup\{U\}=n+1$.
If no confusion arises, we may write $U\cup\{U\}$ as $U+\{*\}$.
The transport along bijections is the one inherited from the species
$F$ but it is restricted to those transformations that keep the
point $*$ fixed.
The stabilizer of a structure $s$ when considered as a $F'-$structure 
is different from its stabilizer as a $F-$structure:
\bd
s\in F'[U]=F[U+\{*\}]~\Rightarrow~H_s^{F'}=H_s^{F}\cap\symm (U).
\ed
As explained in the introduction,
we wish to compare successive levels of a species,
i.e. to compare $F$ with $F'$. In this direction there 
is a small
\begin{lemma}\label{lemma.deriv}
There are only two species (up to multiplicity)
 which satisfy $F=F'$.
\end{lemma}
\noindent\textit{Proof.} Clearly, the species $F$ must have 
the same number of structures at all levels. For $s\in F[n]$, 
the stabiliser $H_s$ satisfies $|H_s|\geq \frac{n!}{|F[n]|}$ 
which for $n$ big enough, reduces the possibilities to either 
the whole symmetric group $\symm (n)$ or the subgroup $A(n)$ 
of even permutations. In the first case we obtain the species 
$E$ of sets which has only one structure at each level, in the 
second we have the species $E^{\pm}$ of \textit{oriented sets}
 with exactly two structures at each level
\bd
E^{\pm}[U]=\{U_+,~ U_-\}
\ed
the stabiliser of each structure being $H_{U_{\pm}}=A(U)$. 

\qed

\noindent
Besides these two ideal cases,
we are interested in species $F$ whose structure at successive levels
``resemble'' each other.
That means that $F_n[U]$ and $F_{n+1}[U+\{*\}]$ should contain structures
that behave similarly under permutations of $U$.
Suppose that we are given a morphism $m$ from a subspecies
$F_1$ of $F'$ to $F$ ($F'=F_1+F_2$).
Then the $F$-structures which belong to the image of this morphism
are similar to their preimages in the sense that their stabilizers
contain those of their preimages.
The action of the morphism $m$ can be encoded in a weight on the species
$F\times F'$.\\

\definitie
A \textit{weighted species} $(F,\om)$ consists of 
\begin{enumerate}
\item
a species of structures $F$
\item
a family of functions $\om_U:F[U]\rightarrow  \C$ called \textit{weights},
\end{enumerate}

\smallskip\noindent
such that for a bijection $\sigma:U\ra V$ one has
$\om_V\circ F[\sigma]=\om_U$.

\smallskip\noindent
The weight $\om_m$ associated to the morphism $m:F_1\rightarrow F$ is
the indicator function of its graph:
\begin{displaymath}
\om_{m,U}(f,g)= \left\{\begin{array}{ll}
\delta_{f,m(g)} & \textrm{if $g\in F_1[U]~,~ f\in F[U]$,} \\
0 & \textrm{if $g\notin F_1[U]$.}
\end{array}\right.
\end{displaymath}
One of the most interesting operations between species is the composition.\\

\definitie
Let $F$ and $G$ two species of structures such that $G[\emptyset]=\emptyset$.
The \textit{composition} $F\circ G$ is a species whose structures on a set
$U$ are made in the following way:
\begin{enumerate}
\smallskip\item
make a partition $\pi$ of the set $U$;
\item
choose an $F$-structure over the set $\pi$: $f\in F[\pi]$;
\item
for each $p\in\pi$ choose a structure $g_p\in G[p]$.
\smallskip
Then the triple $(\pi, f, (g_p)_{p\in\pi})$ is a structure in $F\circ G[U]$.
The transport along $\sigma: U\ra V$ is the natural one.
\end{enumerate}

\noindent
In brief, an $F\circ G$ structure is an $F$-assembly of $G$-structures.
As an example consider the following combinatorial equation.
\be
\mathcal{A}=X\cdot E(\mathcal{A})
\ee
This equation implicitly defines the species $\mathcal{A}$ of rooted trees.
Here is an explicit definition:
\begin{eqnarray}
\mathcal{A}[U]&=&\{f:U\ra U~|
  ~\forall_{u\in U}:f^{\circ k}(u)\hbox{ is eventually constant}\};\nonumber\\
\mathcal{A}[\sigma]&:&f\mapsto\sigma\circ f\circ\sigma^{-1}.\nonumber
\end{eqnarray}
The constant is the root of the tree.
The preimage of the root consists of roots of subtrees.
One can thus consider the tree $f$ as the pair
$(\textrm{root}(f)~, ~\{f_a~|~a\in f^{-1}(\textrm{root}(f))\})$
with $f_a\in \mathcal{A}[U_a]$ the subtree of $f$ with root $a$:
\bd
U_a=\{u\in U~|~\exists k\in\N~\textrm{such that}~f^{\circ k}=a\}\quad,
               \quad f_a=f\upharpoonright_{U_a}
\ed 
We finally note
\bd
U=\{\textrm{root}(f)\}\cup\bigcup_{a\in f^{-1}(\textrm{root}(f))}U_a
\ed
thus completing the bijection between $\mathcal{A}[U]$ and
$X\cdot E(\mathcal{A})[U]$.

\section{Fock Spaces and Analytic Functors}
\label{Secanfunct}

In this section we will describe how one can associate
 to a species of structures an endofunctor of the category 
of Hilbert spaces with contractions. We call the images 
of this functor symmetric spaces associated to the species $F$
 and as we shall see in the following sections, 
they are suitable for constructing algebras of creation and annihilation
 operators, by exploiting the symmetry properties of the species $F$.   

\noindent 
Following Joyal \cite{Joyal.2} we define a special class of
 endofunctors of the category of sets with maps.\\
\definitie
Let $F[\cdot]$ be a species of structures. The \textit{analytic functor}
 $F(\cdot)$ is an endofunctor of the category \textbf{Set} 
of sets with maps, defined by:
\be\label{eq.anfunct} 
F(J)=\tilde{\sum_U} F[U]\times J^U 
\ee

where $J^U=\{c|c:U\ra J\}$ and the symbol $\tilde{\sum}_U$ means 
the set of equivalence classes under bijective transformations:
\bd
F[U]\times J^U\ni(s,c)\mapsto (F[\sigma](s), c\circ \sigma^{-1})
\in F[V]\times J^V
\ed
for $\sigma:U\ra V$. We call the elements of $J$ ``colors''. 
Thus, an element in $F(J)$ is an orbit of J-colored F-structures
 denoted by $[s,c]$. Alternatively
\bd
F(J)=\tilde{\sum_{U}} F[U]\times J^{U} =\sum_{n=0}^{\infty} 
F[n]\times J^{n} ~/~ \symm(n).
\ed

\noindent\textbf{Remark.} This relation can be viewed as a Taylor 
expansion of the set $F(J)$, which explains the name 
``analytic functor'' for $F(\cdot)$ \cite{Joyal.2}. \\

\noindent
Parallel to the functor $F(\cdot)$ we define another endofunctor, 
this time on the category \textbf{Hilb} of Hilbert spaces
 with contractions. 
For any Hilbert space $\mathcal{H}$ and a finite set $U$
 we denote by $\mathcal{H}^{\tens U}$ the Hilbert space arising
 from the positive definite kernel on $\mathcal{H}^{U}$ given by
\bd
k\left( \Tens_{u\in U}\psi_u,\Tens_{u'\in U}\varphi_{u'}\right)=
\prod_{u \in U}\br \psi_u, \varphi_u \ke. 
\ed
For every bijection $\sigma:U\ra V$ there is 
a unitary transformation $U(\sigma):\mathcal{H}^{\tens U}\ra
\mathcal{H}^{\tens V}$ obtained by linear extension of:
\bd
U(\sigma): \Tens_{u\in U}\psi_u \ra \Tens_{v\in V}
\psi_{\sigma^{-1}(v)}.
\ed

\definitie
Let $F$ be a species of structures. For each Hilbert space $\K$
 we construct the \textit{symmetric Hilbert space } 
\be\label{eq.symm.sp}
\G_F(\K):=\Dsum_{n=0}^{\infty} \frac{1}{n!}~
\lsymm (F[n]\rightarrow 
\K^{\tens n})
\ee 

where the subscript ``symm'' denotes the invariance under the natural 
action of the symmetric group $\symm(n)$:
\bd
\Psi \mapsto U(\sigma )\Psi \circ F[\sigma^{-1}]. 
\ed
The factor $\frac{1}{n!}$ refers to the inner product on $\lsymm$.\\

\noindent\textbf{Remark.} There is an equivalent way of writing $\G_F(\K)$:
\be\label{eq.alt.not.symmsp}
\G_F(\K)=\Dsum_{n=0}^{\infty} \frac{1}{n!}~\ell^2 (F[n])
\tens_{\mathrm{S}(n)} \K^{\tens n}
\ee
where $\tens_{\mathrm{S} (n)}$ means that we consider only 
the subspace of the 
tensor product whose vectors are invariant under the action of
 $\symm (n)$.  \\
\noindent
Let us choose an orthonormal basis $(e_j)_{j\in J}$ for the Hilbert
 space $\K$. Let $(e_c)_{c\in J^n}$ be the basis of $\K^{\tens n}$ given by 
$e_c:=\tens_{j\in n}e_{c(j)}$, and 
\begin{eqnarray}
\gamma_{F,J}:F(J)&\ra& [0,\infty)\nonumber \\
\gamma_{F,J}([s,c])&=&|H_{(s,c)}|\nonumber 
\end{eqnarray} 
where $[s,c]$ denotes the orbit of the colored structure $(s,c)$.
\begin{lemma}
There is a unitary equivalence between $\ell^2(F(J),\gamma_{F,J})$
 and $\G_F(\K)$, given by
\bd
U\delta_{[s,c]}=\delta_s\tens_{\mathrm{S}(n)}e_c:=
\sum_{\sigma\in\mathrm{S} (n)}
\delta_{F[\sigma](s)}\tens e_{c\circ\sigma^{-1}}
\ed
\end{lemma}

\noindent\textit{Proof.} Considering $\K^{\tens n}$ as $\ell^2(J^n)$
 we may write 
\begin{eqnarray}
& &U\delta_{[s,c]}=\sum_{\sigma\in\symm (n)}
\delta_{F[\sigma](s)}\tens e_{c\circ\sigma^{-1}}   \nonumber\\
& &=\sum_{\sigma\in\symm (n)}\delta_{F[\sigma](s),c\circ\sigma^{-1}}
=|H_{(s,c)}|\cdot \mathbf{1}_{[s,c]}                            \nonumber
\end{eqnarray}

\noindent
It follows that 
\begin{eqnarray}
\|U\delta_{[s,c]}\|^2&=&\frac{1}{n!}|H_{(s,c)}|^2\cdot
|[s,c]|=|H_{(s,c)}|        \nonumber\\
&=&\|\delta_{[s,c]}\|^2.  \nonumber
\end{eqnarray}
Since the functions $\mathbf{1}_{[s,c]}$ span the space
 $\ell^{2}_{\mathrm{symm}}(F[n]\times J^n)$, the operator
 $U$ is surjective and hence unitary.

\qed

\noindent 
\textbf{Remark.} For a constant coloring $c$ we have
 $\|\delta_{[s,c]}\|^2=|H_s|$, whereas for all colors different,
 $\| \delta_{[s,c]}\|^2=1$.\\
\noindent

Certain operations with species of structures extend to 
analytic species \cite{Joyal.2} and to the symmetric spaces:
 addition, multiplication and substitution. \\

\noindent\textbf{1) Addition.} As $(F+G)(J)$ is the disjoint union 
of $F(J)$ and $G(J)$, we have 
\bd
\G_{F+G}(\K)=\G_F(\K)\oplus \G_G(\K).
\ed

\noindent\textbf{2) Multiplication.} Similarly, we have
\begin{eqnarray}
& & (F\cdot G)(J)  = \tilde{\sum_U}(\sum_{U_1+U_2=U}F[U_1]
\times G[U_2])\times J^U                                    \nonumber \\
 & = & \tilde{\sum_U}\sum_{U_1+U_2=U}F[U_1]\times G[U_2]
\times J^{U_1+U_2}  =
\tilde{\sum_{U_1,U_2}}F[U_1]\times G[U_2]\times J^{U_1+U_2}  \nonumber \\
& = & F(J)\times G(J).\nonumber
\end{eqnarray}
which suggests the following unitary transformation from 
$\G_{F\cdot G}(\K)$ to \\ 
$\G_F(\K)\tens\G_G(\K)$:
\bd
T: \delta_{[(f,g), c]}\ra \delta_{[f,c_1]}\tens\delta_{[g,c_2]}
\ed
for $f\in F[n]$, $g\in G[m]$, $c\in J^{m+n}$ 
and $c_1,c_2$ the restrictions of $c$ to $n$ respectively 
$m$. Indeed the map preserves orthogonality and is isometric:
\begin{eqnarray}
\|\delta_{[(f,g), c]}\|^2 & = &  |H_{((f,g),c)}| 
  =  |H_{(f,c_1)}\cdot H_{(g,c_2)}|
=  |H_{(f,c_1)}|\cdot | H_{(g,c_2)}|\nonumber \\
 & = & \|\delta_{[f,c_1]}\|^2\cdot \|\delta_{[g,c_2]}\|^2 \nonumber
\end{eqnarray}
From now on we will consider  $\G_{F\cdot G}(\K)$ and 
$\G_F(\K)\tens\G_G(\K)$ as identical, without mentioning the unitary $T$.\\

\noindent\textbf{3) Substitution.} we start with the analytic functors:
\begin{eqnarray}
(F\circ G)(J) & = & \tilde{\sum_U}(F\circ G)[U]\times J^U \nonumber \\
=\tilde{\sum_U}(\tilde{\sum_\pi}F[\pi]\times G^\pi[U])\times J^U & = &
\tilde{\sum_\pi}(F[\pi]\times \tilde{\sum_U} G^\pi[U])\times J^U \nonumber \\
=\tilde{\sum_\pi}F[\pi]\times G^\pi(J) & = & F(G(J))\nonumber
\end{eqnarray}
where we have used $G^\pi(J)=G(J)^\pi$, which follows from the
 multiplication property. At the level of symmetric spaces we have
 the unitary transformation from $\G_F(\G_G(\K))$ to $\G_{F\circ G}(\K)$:
\bd
T: \delta_{[f,C]}\ra \delta_{[f,(g_a)_{a\in \pi}, c]}
\ed
with the following relations for the structures appearing above: 
$f\in F[\pi]$, $C:\pi\ra G(J)$ such that $C(a)=[g_a, c_a]$, and 
$c\upharpoonright_a=c_a$. Let us check the isometric property:
\begin{eqnarray} 
\|\delta_{[f,C]}\|^2 & = & \prod_{a\in \pi}\|\delta_{C(a)}\|^2\cdot
 |H_{(f,C)}|
= \prod_{a\in \pi}\|\delta _{[g_a,c_a]}\|^2\cdot |H_{f,C}|  \nonumber \\
& = & \prod_{a\in \pi}|H_{(g_a,c_a)}|\cdot |H_{f,C}|
 = \|\delta_{[f,(g_a)_{a\in \pi}, c]}\|^2 \nonumber
\end{eqnarray}

\noindent\textbf{Symmetric Fock space.} The symmetric Hilbert space
 associated to the species of sets is the well known symmetric Fock space:
\bd
\G_E(\K)=\Dsum_{n=0}^\infty\frac{1}{n!}\lsymm(E[n]\ra
 \K^{\tens n})= \Dsum_{n=0}^\infty\frac{1}{n!}\K^{\tens_{s} n}.  
\ed
\noindent\textbf{Full Fock space.} For the linear orders we obtain
 the full Fock space:
\bd
\G_L(\K)=\Dsum_{n=0}^\infty\frac{1}{n!}\lsymm(L[n]\ra \K^{\tens n})= \Dsum_{n=0}^\infty\K^{\tens {n}}.  
\ed
\noindent\textbf{Antisymmetric Fock space.} We recall from lemma 
\ref{lemma.deriv} that the species  $E^{\pm}$ of oriented sets has 
two structures at all levels 
\bd
E^{\pm}[U]=\{U_+,~U_-\}
\ed
which are mapped into each other by odd permutations and have 
as stabiliser the group $A(U)$ of even permutation. The representation 
of $\symm (n)$ on $\ell^2(E^{\pm}[n])$ contains two 
one-dimensional irreducible sub-representations, the symmetric and 
the antisymmetric representation. Accordingly the symmetric Hilbert 
space associated to $E^{\pm}$ is the direct sum of the symmetric and 
antisymmetric Fock spaces:
\bd
\G_{E^{\pm}}(\K)=  \G_{s}(\K)\oplus \G_{a}(\K). 
\ed
\noindent\textbf{Remark.} The set of species as defined in the previous
 section can be enlarged by defining \cite{Ber.Lab.Ler.} the 
\textit{virtual species} as equivalence classes of pair of species
 of structures under the equivalence relation:
\bd
(F_1,G_1)\thicksim (F_2,G_2)\Leftrightarrow F_1+G_2=F_2+G_1
\ed
One denotes the equivalence class of $(F,G)$ by $F-G$. Thus we can
 say that the antisymmetric Fock space is associated to the virtual
 species $E^{\pm}-E$.

\section{Creation and Annihilation Operators}
\label{Secalgebras}
In this section we will describe a general framework for constructing 
*-algebras of operators on symmetric Hilbert spaces by giving the 
action of the generators of these algebras, the creation and 
annihilation operators. In particular in the case of the species of
 sets $E$ and linear orderings $L$, we obtain the well known 
canonical commutation relations algebra (C.C.R.), respectively 
the algebra of creation/annihilation operators on the full Fock space.\\
The starting point is the observation that the operation of 
derivation of species of structures can be interpreted as 
removal of  point $*$ from a structure. This makes it possible 
to define operators between the symmetric Hilbert spaces of a 
species of structure $F$ and its derivative $F'$. \\
\noindent
We will consider now ``colored'' $F$-structures. Let $J$ be the 
set of ``colors'' and $i\in J$. We have the map
\begin{displaymath}
a^{*}(i):F'[U]\times J^{U}\ra  F[U+\{*\}]\times J^{U+\{*\}}
\end{displaymath}
such that
\bd
a^{*}(i): (s,c)\ra (s, c^{+}_{i})
\ed
where $c^{+}_{i}:U+\{*\}\ra J$ is given by:
\bd
c^{+}_{i}(u)=\left\{\begin{array}{ll}
c(u) & \textrm{if $u\in U$} \\
 i & \textrm{if $u=*$}
\end{array}\right.
\ed
\noindent
As we did in section \ref{Secanfunct}, we pass to the set of orbits 
of $J$-colored $F$-structures. The map $a^{*}(i)$ projects to a 
well defined map from $F'(J)$ to $F(J)$:
\begin{eqnarray}
a^{*}(i):F'[U]\times J^{U}/\symm (U) & \ra & F[U+\{*\}]
\times J^{U+\{*\}}/\symm (U+\{*\})                           \nonumber \\
                     a^{*}(i): [s,c] & \ra & [s,c^{+}_{i}]   \nonumber    
\end{eqnarray}
\noindent
But as $F(J)$ determines an orthogonal basis of the space $\G_F (\K)$ 
for $(e_j)_{j\in J}$ orthogonal basis in $\K$, we can  extend  
$a^{*}(i)$ by linearity to an operator
\begin{displaymath} 
a^{*}(i):\G_{F'}(\K)\ra \G_F(\K).
\end{displaymath}
\noindent
The adjoint of $ a^{*}(i)$ acts in the opposite direction:
\begin{displaymath}
a(i):\G_F(\K)\ra \G_{F'}(\K).
\end{displaymath}

The problem with this definition is that in general the species
 $F$ and $F'$ are distinct which means that one cannot take the 
``field operators'' $a^*(i)+a(i)$ and only certain products of 
creation and annihilation operators are well defined. In section
 \ref{Secspecies} we pointed out that the ``similarity'' of the 
structures of the species $F$ and $F'$ can be encoded in a 
weight on the cartesian product $F\times F'$. Let $\om$ be 
such a weight. Then $\om_{U}: F[U]\times F'[U]\ra \C$ such 
that for all $s\in F[U], ~t\in F'[U]$ and $\sigma :U \ra W$ we have:
\begin{displaymath}
\om (s,t)= \om (F[\sigma](s), F'[\sigma](t))
\end{displaymath} 
\noindent
We will use this to define creation and annihilation operators 
which act on the same space $\G_F(\K)$. In the sequel we will
 refer to the pair $(\G_F(\K)~,~ \om)$ as 
\textit{combinatorial Fock space}.\\
\definitie
a) The \textit{annihilation operator} (before symmetrization) 
associated to the species $F$ and weight $\om$ is defined by:
\begin{displaymath}
\tilde{a}(h):\Dsum_{n=0}^{\infty}\frac{1}{n!}\ell^{2} (F[n]\ra 
\K^{\tens n})\ra \Dsum_{n=0}^{\infty}\frac{1}{n!}\ell^{2} (F[n]
\ra \K^{\tens n})
\end{displaymath}
\begin{displaymath}(\tilde{a}(h)\varphi)(f)=\sum_{g\in 
F[n+1]}\om (f,g)\cdot \mathrm{inp}_{n}(h,\varphi (g))
\end{displaymath}
\noindent
where $f\in F[n]~,~h\in\K$ and $\mathrm{inp}_{k}(h,\cdot)$ 
is the operator:
\begin{displaymath}
\mathrm{inp}_{k}(h,\psi_0\tens ..\tens\psi_n)=\br h, \psi_{k}
\ke\psi_0\tens ..\tens\psi_{k-1}\tens\psi_{k+1}\tens ..\tens\psi_n
\end{displaymath}
for $k\in \{0, 1,..,n\}$.
\newline
b) The \textit{creation operator} (before symmetrization) is:
\begin{displaymath}
\tilde{a}^*(h):\Dsum_{n=0}^{\infty}\frac{1}{n!}\ell^{2} (F[n]\ra 
\K^{\tens n})\ra \Dsum_{n=0}^{\infty}\frac{1}{n!}\ell^{2} (F[n]\ra 
\K^{\tens n})
\end{displaymath}  
\begin{displaymath}
(\tilde{a}^*(h)\varphi)(f)=(n+1)\cdot\sum_{g\in F[n]}
\overline{\om (g,f)}\cdot \mathrm{tens}_{n}(h, \varphi (g))
\end{displaymath}
where $f\in F[n+1]~,~h\in \K$ and $\mathrm{tens}_{k}(h,\cdot)$ 
is the operator:
\begin{displaymath}
\mathrm{tens}_{k}(h,\psi_0\tens..\tens\psi_{n-1})=
\psi_0\tens..\tens\psi_{k-1}\tens h\tens\psi_{k}\tens..\tens\psi_{n-1}
\end{displaymath}
for $k\in \{0,1,..,n\}$.\\
\textbf{Remark.} In order to avoid domain problems for 
$\tilde{a}, \tilde{a}^*$, we will restrict to weights which are bounded,
$|\omega (t,s)|\leq C$ for all $t,s$. Then
\bd
\|a(h)\psi_n\|\leq n^{\frac{1}{2}}C\|h\|\|\psi_n\|~~~~
\|a^*(h)\psi_n\|\leq (n+1)^{\frac{1}{2}}C\|h\|\|\psi_n\|
\ed 
for $\psi_n\in\ell^{2} (F[n]\ra\K^{\tens n})$ 
thus $a(h), a^*(h)$ have well defined extensions to the 
domain $D(N^{\frac{1}{2}})$ $(N\psi_n=n\psi_n)$. As this will not play 
a major role here, we will omit specifying the domain, usually the 
the vectors considered should belong to $D(N^{\frac{1}{2}})$. \\
\noindent
We consider now the symmetrized creation and annihilation operators 
which act on the symmetric Hilbert space and which are the main 
object of our investigation.
\begin{lemma}\label{ann}
The unsymmetrized annihilation operator $\tilde{a}(h)$ restricts
 to a well defined operator $a(h)$ on the symmetric Hilbert space 
$\G_F(\K)$:
\end{lemma}

\textit{Proof.} Let $\varphi\in\G_F(\K)$. Then 
$\varphi(F[\sigma ](f))=U(\sigma )\varphi (f)$ for all 
$\sigma \in \symm (n )~,~ \\ 
f\in F[n ]$ and 
\begin{eqnarray}
(a(h)\varphi)(F[\sigma](f)) & = &
\sum_{g}\om(F[\sigma](f), g)\cdot \mathrm{inp}_n(h,\varphi(g))\nonumber \\
 =\sum_{g'}\om(f, g')\cdot \mathrm{inp}_n(h,\varphi(F[\tilde{\sigma}]g'))
 & = & \sum_{g'}\om(f, g')\cdot 
U(\sigma)\mathrm{inp}_n(h,\varphi (g'))\nonumber \\
 &= & U(\sigma)(a(h)\varphi)(f)\nonumber
\end{eqnarray}
where $\tilde{\sigma}:n+1\ra n+1$ is given by
\begin{displaymath}
\tilde{\sigma}(i)=\left\{\begin{array}{ll}
\sigma (i) & \textrm{if $i\in n$} \\
 n & \textrm{if $i=n$}
\end{array}\right.
\end{displaymath}
\qed

\begin{lemma}
The operator $\tilde{a}^*(h)$ is the adjoint of $\tilde{a}(h)$ 
on the unsymmetrized space 
$\Dsum_{n=0}^\infty\frac{1}{n!}\ell^2(F[n]\ra \K^{\tens n})$.  
\end{lemma}
\textit{Proof.} Let $\varphi, \psi$ be two vectors in 
$\Dsum_{n=0}^\infty\frac{1}{n!}\ell^2(F[n]\ra \K^{\tens n})$ 
and  $\varphi_n, \psi_n \in\frac{1}{n!}\ell^{2}(F[n]\ra\K^{\tens n})$
their components on level $n$.\\
\noindent
Then we have:
\begin{eqnarray}
& & \br\psi, \tilde{a}^*(h)\varphi\ke
=\sum_{n=0}^\infty\frac{1}{(n+1)!}\sum_{g}(n+1)\br\psi_{n+1}(g),
 (\tilde{a}^*(h)\varphi_n)(g)\ke                    \nonumber \\
& = & \sum_{n=0}^\infty\frac{1}{n!}\sum_{f,g}\br\psi_{n+1}(g), 
\overline{\om(f,g)}\cdot\varphi_n(f)\tens h\ke              \nonumber \\
& = & \sum_{n=0}^\infty\frac{1}{n!}\sum_{f,g}
\br\om(f,g)\cdot\mathrm{inp}_n(h,\psi_{n+1}(g)), 
\varphi_n(f)\ke      \nonumber\\
& = & \sum_{n=0}^\infty\frac{1}{n!}\sum_{f}
\br(\tilde{a}(h)\psi_{n+1})(f), \varphi_n(f)\ke
= \br\tilde{a}(h)\psi, \varphi\ke \nonumber
\end{eqnarray}
\qed

We will restrict our attention to the action of the creation and 
annihilation operators on the symmetric Hilbert space $\G_F(\K)$. 
From Lemma \ref{ann} the annihilation operator $a(h)$ is well defined
 on  $\G_F(\K)$. We call its adjoint  on the symmetric Hilbert space,
 the symmetrized creation operator. If $P$ is the projection
 to $\G_F(\K)$  then the symmetrized creation operator is:
\begin{displaymath}
a^*(h)\varphi=P\tilde{a}^*(h)\varphi \textrm{ for } \varphi \in  \G_F(\K)
\end{displaymath}

\begin{lemma}
Let $f\in F[n+1]$ and $\tau_{n,k}\in\symm (n+1)$
 the transposition of $n$ and $k$. 
 Then for any $\varphi\in \G_F(\K)$ the action 
of the symmetrized creation operator has the expression:

\begin{displaymath}
(a^*(h)\varphi)(f)=\sum_{k=0}^{n}\sum_{g}
\overline{\om(g, F[\tau_{n,k}](f))}
\cdot U(\tau_{n,k})(\varphi(g)\tens h).
\end{displaymath}

\end{lemma}
\textit{Proof.} We have:

\begin{eqnarray}
 & &(P\tilde{a}^*(h)\varphi)(f)=
\frac{1}{(n+1)!}\sum_{\sigma \in \symm(n+1)}
U(\sigma)(\tilde{a}^*(h)\varphi)(F[\sigma^{-1}]f)\nonumber \\
\label{sum}
 & = & \frac{1}{n!}\sum_{\sigma \in \symm(n+1)}\sum_{g}
\overline{\om(g,F[\sigma](f))}\cdot U(\sigma^{-1})(\varphi(g)\tens h)
\end{eqnarray}

If $\sigma\in\symm(n+1)$ and  $\sigma^{-1}(n)=k$ then 
$\rho=\tau_{n,k}\circ\sigma^{-1}\in\symm(n)$. Thus the 
sum over all permutations can be split into a sum over 
$k\in n+1$ and one over $\symm(n)$. Moreover, 
from the definitions of  $\G_F(\K)$ and that of a weight we know that 
\begin{eqnarray}
U(\rho)\varphi(g) & = & 
\varphi(F[\rho](g))\nonumber \\ 
\om(g,F[\rho^{-1}\circ\tau_{n,k}](f)) & = &
\om(F[\rho](g),F[\tau_{n,k}](f))\nonumber 
\end{eqnarray} 
which substituted into the sum (\ref{sum}) gives:
\begin{eqnarray}
& & \frac{1}{n!}\sum_{k=0}^{n}\sum_{\rho\in \symm (n)}\sum_{g}
\overline{\om(F[\rho](g),F[\tau_{n,k}](f))}\cdot 
U(\tau_{n,k})(\varphi(F[\rho](g))\tens h)
\nonumber \\
& = & \sum_{k=0}^{n}\sum_{g'}
\overline{\om(g',F[\tau_{n,k}](f))}
\cdot U(\tau_{n,k})(\varphi(g')\tens h).\nonumber
\end{eqnarray}

\qed 

\noindent
Sometimes algebras are defined by giving relations among generators 
as for example commutation relations. We will give next explicit 
formulas for the product of a creation and an annihilation 
operator.
\begin{lemma}\label{a*a}
Let  $f\in F[n]$  and $\varphi\in \G_F(\K)$. Then
\begin{equation}
(a^*(h_1)a(h_2)\varphi)(f)=\sum_{k=0}^{n-1}
\sum_{f'}(\overline{\om}\cdot\om)_{k}(f,f')\cdot
\textrm{tens}_k(h_1, \textrm{inp}_k(h_2, \varphi (f')))
\end{equation}
where we have made the notation
\begin{equation}\label{low.neigh}
(\overline{\om}\cdot\om)_{k}(f,f')=
\sum_{g}\overline{\om(g, F[\tau_{n-1,k}](f))}\om(g,F[\tau_{n-1,k}](f'))
\end{equation}

\end{lemma}   
\noindent\textit{Proof.} By applying successively 
the definitions of $a^*(h_1)$ and $a(h_2)$ we have:
\begin{eqnarray} 
& & (a^*(h_1)a(h_2)\varphi)(f)
=\sum_{k=0}^{n-1}\sum_{g}\overline{\om(g, F[\tau_{n-1,k}](f))}\cdot 
U(\tau_{n-1,k})(a(h_2)\varphi)(g)\tens h_1 \nonumber \\
& & =\sum_{k=0}^{n-1}\sum_{g,f'}\overline{\om(g, F[\tau_{n-1,k}](f))}
\om(g,f')\cdot U(\tau_{n-1,k})(\textrm{inp}_{n-1}(h_2,\varphi(f'))\tens h_1 )                                   \nonumber \\
& &= \sum_{k=0}^{n-1}\sum_{g,f'}\overline{\om(g, F[\tau_{n-1,k}](f))}
\om(g,F[\tau_{n-1,k}]f')\cdot
\textrm{tens}_k(h_1,\textrm{inp}_k(h_2,\varphi(f')))\nonumber \\
& &= \sum_{k=0}^{n-1}\sum_{f'}(\overline{\om}\cdot\om)_{k}(f,f')\cdot
\textrm{tens}_k(h_1, \textrm{inp}_k(h_2, \varphi (f'))).\nonumber
\end{eqnarray}

\qed

\begin{lemma}\label{aa*}
Let $f\in F[n]$ and $\varphi\in \G_F(\K)$. Then:
\begin{eqnarray}
(a(h_1)a^*(h_2)\varphi)(f) & = & \sum_{k=0}^{n-1}
\sum_{f'}(\om\cdot\overline{\om})_k(f,f')\cdot
\mathrm{tens}_k(h_2, \mathrm{inp}_k(h_1, \varphi (f')))\nonumber \\
& + &\br h_1,h_2\ke\cdot
\sum_{f'} (\om\cdot\overline{\om})_n(f,f')\varphi(f')   \nonumber
\end{eqnarray}
where we have made the notation
\begin{equation}\label{up.neigh}
(\om\cdot\overline{\om})_k(f,f')=\sum_{g}\om(f,g)
\overline{\om(f',F[\tau_{n,k}](g))}
\end{equation}

\end{lemma}

\noindent\textit{Proof.} We use the definitions 
of $a(h_1)$ and $a^*(h_2)$:
\begin{eqnarray}
& & (a(h_1)a^*(h_2)\varphi)(f)
=\sum_{g}\om(f,g)\cdot\textrm{inp}_n(h_1,(a^*(h_2)\varphi)(g))\nonumber \\
& = & \sum_{g,f'}\sum_{k=0}^{n}\om(f,g)
\overline{\om(f',F[\tau_{n,k}](g))}
\cdot\textrm{inp}_n(h_1,U(\tau_{n,k})(\varphi(f')\tens h_2))
\nonumber \\
& = & \sum_{k=0}^{n-1}\sum_{f'}(\om\cdot\overline{\om})_k(f,f')
\cdot\textrm{tens}_k(h_2, \textrm{inp}_k(h_1, \varphi (f')))  \nonumber \\
& + &\br h_1,h_2\ke\cdot
\sum_{f'} (\om\cdot\overline{\om})_n(f,f')\varphi(f')  \nonumber
\end{eqnarray}

\qed

\subsection{Examples.} We will describe a few known operator 
algebras in the language developed so far and a new algebra 
based on the species $\mathcal{A}$ of rooted trees.\\
 
\textbf{1) Sets:}  The combinatorial Fock space is $(E, \om_E)$ 
with $E[U]=\{U\}$ and $\om(\{U\},\{U+\{*\}\})=1$. We use lemmas
 \ref{a*a} and \ref{aa*} to calculate the commutator of the
 creation and annihilation operator:
\begin{eqnarray}
& & (a(h_1)a^*(h_2)-a^*(h_2)a(h_1))\varphi(f)
=\br h_1, h_2\ke\cdot 
(\om\cdot\overline{\om})_n(f,f')\varphi(f')\nonumber \\
& + & \sum_{k=0}^{n-1}((\om\cdot\overline{\om})_k(f,f')-
(\overline{\om}\cdot\om)_k(f,f'))
\cdot\textrm{tens}_k(h_2, \textrm{inp}_k(h_1, \varphi(f'))\nonumber 
\end{eqnarray}
\noindent
But $(\om\cdot\overline{\om})_k(f,f')=
(\overline{\om}\cdot\om)_k(f,f')=
(\om\cdot\overline{\om})_n(f,f')=\delta_{f,f'}$ for all $k\in n$ 
which implies the C.C.R.:
\bd
a(h_1)a^*(h_2)-a^*(h_2)a(h_1)=\br h_1, h_2\ke\mathbf{1}
\ed
\noindent
In particular it is clear that $\G_E(\K)$ is the symmetric 
Fock space over the Hilbert space $\K$.\\

\textbf{2) Linear Orders:} Let $(L,\om_L)$ be the combinatorial 
Fock space with\bd
L[U]=\{f:U\ra\{0,1,..,|U|-1\}\}
\ed
and
\begin{displaymath}
\om_L(f,g)=\delta_{f,g\upharpoonright_{U}} ~\textrm{for}~ 
f\in L[U],~ g\in L[U+\{*\}]
\end{displaymath}
where 
\begin{displaymath}
\delta_{f,g\upharpoonright_{U}}=\left\{\begin{array}{ll}
1 & \textrm{if $f(u)=g(u)$ for $u\in U$}  \\
0 & \textrm{otherwise}
\end{array}\right.
\end{displaymath}
\noindent
From (\ref{up.neigh}) we have:
\bd
(\om\cdot\overline{\om})_k(f,f')=\sum_{g}\delta_{f,g\upharpoonright_{U}}
\cdot\delta_{f',L[\tau_{n,k}](g)\upharpoonright_{U}}=
\delta_{k,n}\cdot\delta_{f,f'}
\ed
\noindent
Then by applying Lemma \ref{aa*} we obtain
\bd
a(h_1)a^*(h_2)=\br h_1,h_2\ke\mathbf{1}
\ed
\noindent
which characterizes the algebra of creation and annihilation 
operators on the full Fock space \cite{Voi.Dy.Ni.}. \\
 
\textbf{3) Oriented Sets:} We refer to the previous sections for 
the definition of the species $E^{\pm}$ of oriented sets. 
The weight $\om_{E^{\pm}}$ is given by
\begin{eqnarray}
\om_{E^{\pm}}(U_{+},U^*_{+}) & = & \om_{E^{\pm}}(U_{-},U^*_{-}) =  1,
\nonumber \\
\om_{E^{\pm}}(U_{+},U^*_{-}) & = & \om_{E^{\pm}}(U_{-},U^*_{+}) =  0 
\end{eqnarray}
\noindent
where $U^*=U+\{*\}$.
With the help of the ``switching'' sign operator 
\bd
\mathbf{g}\varphi(\pm)=\varphi(\mp)
\ed 
\noindent
we obtain the $\mathbf{g}$-commutation relations:
\be\label{gcomm}
a(h_1)a^*(h_2)-\mathbf{g}a^*(h_2)a(h_1)=\br h_1, h_2\ke \mathbf{1}
\ee
As we saw in the previous section, the space $\G_{E^{\pm}}(\K)$ 
is isomorphic to the direct sum of the symmetric and antisymmetric 
Fock space over $\K$:
\bd
\G_{E^{\pm}}(\K)=\G_s(\K)\oplus\G_a(\K)
\ed   
through the transformation:
\bd
\varphi_s=\varphi(+)+\varphi(-) ~~\textrm{and}~~\varphi_a=
\varphi(+)-\varphi(-) 
\ed
then the $\mathbf{g}$-commutation relations can be
 written equivalently as: 
\bd
(a(h_1)a^*(h_2)-a^*(h_2)a(h_1))\varphi_s=\br h_1, h_2\ke\varphi_s
\ed
and
\bd
(a(h_1)a^*(h_2)+a^*(h_2)a(h_1))\varphi_a=\br h_1, h_2\ke\varphi_a
\ed

\textbf{4) Rooted Trees:} We recall the definition of the species
 $\mathcal{A}$: 
\bd
\mathcal{A}[U]=\{f:U\ra U~|~f^{\circ k}(u)=
\textrm{root}(f)\in U~\textrm{for}~k\geq |U|, u\in U\}
\ed
\noindent
with the transport along $\sigma$: $\mathcal{A}[\sigma](f)
=\sigma\circ f \circ \sigma^{-1}$. We note that 
$\mathcal{A}[\emptyset]=\emptyset$. We consider a natural weight 
which can be described as follows: it takes value 1 on those 
pairs of trees for which the second is obtained by adding a leaf 
to the first one, and takes value 0 for the rest. Thus for 
$t_1\in \mathcal{A}[U]$ and  $t_2\in \mathcal{A}[U+\{*\}]$ the weight is:
\begin{displaymath}
\om_\mathcal{A}(t_1,t_2)=
\left\{\begin{array}{ll}
1 & \textrm{if $t_1(u)=t_2(u)$ for $u\in U$} \\
 0 & \textrm{otherwise}
\end{array}\right\}
:=\delta_{t_1, t_2\upharpoonright_U}
\end{displaymath}
We will compute the commutator of $a(h_2)$ with $a^*(h_1)$. 
For this we need to obtain the expressions of 
$(\overline{\om}\cdot\om)_k(\cdot,\cdot)$ 
and $(\om\cdot\overline{\om})_k(\cdot,\cdot)$. We start with

\begin{eqnarray}
(\om\cdot\overline{\om})_n(f,f') 
 & = & \sum_{g}\om_\A(f,g)\cdot\overline{\om_\A(f',g)} 
   =  \sum_{g}\delta_{f, g\upharpoonright_n}
\cdot\delta_{f', g\upharpoonright_n} \nonumber \\
 & = & \delta_{f,f'}\sum_{g}\delta_{f, g\upharpoonright_n}
   =  n\delta_{f,f'} \label{trees0}
\end{eqnarray}
The factor $n$ appears because there are $n$ possible way of
 attaching a leaf to the tree $f$ each one giving a tree $g$ such
 that $g\upharpoonright_n=f$. For $k<n$ we have
\bd
(\om\cdot\overline{\om})_k(t,t')=\sum_{g}\om_\A(t,g)\cdot
\overline{\om_\A(t',\A[\tau_{n,k}](g))}=
\sum_{g}\delta_{t, g\upharpoonright_n}
\cdot\delta_{t',\A[\tau_{n,k}](g)\upharpoonright_n}.
\ed
At most one term in this sum is different from zero, for the tree
 $g$ satisfying:
\begin{equation}\label{trees1}
\left\{\begin{array}{ll}
g(i)=t(i) & \textrm{if $i\in n$} \\
g(j)=t'(j)& \textrm{if $j\in n\setminus\{k\}$}\\
g(n)=t'(k)&
\end{array}\right.
\end{equation}
\noindent
On the other hand 
\bd
(\overline{\om}\cdot\om)_k(t,t')
=\sum_{g'}\overline{\om_\A(g',\A[\tau_{n-1,k}](t))}
\cdot\om_\A(g',\A[\tau_{n-1,k}](t'))
\ed
\bd
=\sum_{g}\delta_{g',\A[\tau_{n-1,k}](t) 
\upharpoonright_{n-1}}
\cdot \delta_{g',\A[\tau_{n-1,k}](t') 
\upharpoonright_{n-1}}
=\delta_{\A[\tau_{n-1,k}](t) 
\upharpoonright_{n-1},~\A[\tau_{n-1,k}](t') 
\upharpoonright_{n-1}}
\ed
\begin{equation}\label{trees2}
=\left\{\begin{array}{ll}
1 & \textrm{if $t(i)=t'(i)$ for all $i\in n, i\neq k$} \\
0 & \textrm{otherwise}
\end{array}\right.
\end{equation}
Finally from (\ref{trees1}), (\ref{trees2}) we conclude that 
$(\overline{\om}\cdot\om)_k(t,t')=(\om\cdot\overline{\om})_k(t,t')$ 
for $k\in\{0,1,..n-1\}$.\\
\noindent
Let us define the ``vertex number'' operator $N$ by
\bd
 (N\varphi)(t)=n\varphi(t)
\ed
for $t\in \A[n]$. The usual commutation relations 
between $N$ and the creation operator hold:
\bd
[N, a^*(h)]=a^*(h)
\ed
By using Lemmas \ref{a*a} and \ref{aa*} we obtain the following:
\begin{theorem}\label{th.rootedtrees}
The following commutation relations hold on the combinatorial 
Fock space $(\A,\om_\A)$:
\be
a(h_1)a^*(h_2)-a^*(h_2)a(h_1)=N\br h_1,h_2\ke
\ee
\end{theorem}

\noindent\textbf{Remark:} Notice that the vacuum of $\G_\A(\K)$ 
is an eigenvector of $N$ with eigenvalue 1.\\

\textbf{5) Simple Directed Graphs:} Let us define a species whose 
structures are directed graphs for which any pair of vertices is 
connected by at most one edge:
\be
\D[U]=\{g\in U\times U~|~(u,v)\in g \Rightarrow (v,u)\notin g \}
\ee 
where the transport along $\sigma$ is given by $\sigma\times\sigma$.\\
\noindent
Let $g_1\in \D[U]$ and $g_2\in \D[U+\{*\}]$. Then $\om(g_1, g_2)\neq 0$ 
if and only if $g_2$ contains $g_1$ as a subset and all edges of $g_2$
 connecting the vertex $*$ with vertices in $U$ are oriented from
 $*$ to $U$. We make the following convenient notation for the set
 of edges going out of a vertex $a$ of $g\in \D[V]$  :
\bd
v_a(g)=\{(a,v)~|~ (a,v)\in g\}=\{ a\}\times e_a(g)
\ed
The weight $\om^{\D ,q}$ depends on the real parameter 
$0\leq q\leq 1$  and is defined by:

\bd
\om^{\D ,q}(g_1,g_2)=\delta_{g_2~,~g_1+v_*(g_2)}
\cdot(q^{|U|-|v_*(g_2)|}\cdot(1-q)^{|v_*(g_2)|})^\frac{1}{2}
\ed
In the rest of this section we prove that $(\D, \om^{\D ,q})$  
is a realization of the q-commutation relations
 (\cite{Grb., Fiv., Zag., Boz.Sp.2, Maa.vLee.}).
\begin{theorem}
On  $(\D, \om^{\D ,q})$ we have:
\bd
a(h_1)a^*(h_2)-q\cdot a^*(h_2)a(h_1)=\br h_1, h_2\ke
\ed 
\end{theorem} 
\textit{Proof.} We employ lemmas \ref{aa*} and \ref{a*a}. First:
\begin{eqnarray}
 & & (\om\cdot\overline{\om})_n(f,f')=\sum_{g}\om^{\D ,q}(f,g)\cdot
\overline{\om^{\D ,q,}(f',g)} \nonumber \\
 & & =\sum_{g}\delta_{g, f+v_n(g)}\cdot
\delta_{g, f'+v_n(g)}\cdot q^{n-|v_n(g)|}\cdot
(1-q)^{|v_n(g)|}\nonumber \\
 & & =\delta_{f,f'}\cdot\sum_{v_n\subset n}q^{n-|v_n|}\cdot
(1-q)^{|v_n|}=\delta_{f,f'}
\end{eqnarray}
\noindent
It remains to be proved that $(\om\cdot\overline{\om})_k(f,f')=
q\cdot (\overline{\om}\cdot\om)_k(f,f')$ for \\
$k\in\{0,1,..n-1\}$ and $f,f'\in\D[n]$.\\
\noindent
In the sum
\bd
(\om\cdot\overline{\om})_k(f,f')=\sum_{g}\om^{\D ,q}(f,g)\cdot
\overline{\om^{\D ,q,}(f',\D[\tau_{n,k}](g))}
\ed
the only nonzero contribution comes from $g\in \D[U+\{*\}]$ such that:
\bd
g=f+\{n\}\times e_n(g)=(f\setminus v_k(f))+v_k(f)+\{n\}\times e_n(g)
\ed
and
\begin{eqnarray}
\D[\tau_{n,k}](g) & = & (f\setminus v_k(f))+\{n\}
\times e_k(f)+\{ k\}\times e_n(g) \nonumber \\
                  & =  & f'+\{n\}\times e_k(f) \nonumber
\end{eqnarray}
which together imply
\bd
(f\setminus v_k(f))+\{k\}\times e_n(g)=f'.
\ed
But this means that $e_n(g)=e_k(f')$ and 
$f\setminus v_k(f)=f'\setminus v_k(f')$. Then we have the expression
\be\label{graphsup.neigh}
(\om\cdot\overline{\om})_k(f,f')=\delta_{f\setminus v_k(f),f'
\setminus v_k(f')}\cdot
 q^{n-\frac{|v_k(f)|+|v_k(f')|}{2}}\cdot(1-q)^{\frac{|v_k(f)|+|v_k(f')|}{2}}
\ee
On the other hand in $(\overline{\om}\cdot\om)_k(f,f')$ 
we get only the contribution from 
those $g$ for which:
\bd
g'=\D[\tau_{n-1,k}](f)\upharpoonright_{n-1} =
\D[\tau_{n-1,k}](f')\upharpoonright_{n-1}
\ed
Thus we obtain
\be\label{graphslow.neigh}
(\overline{\om}\cdot\om)_k(f,f')=
\delta_{f\setminus v_k(f),f'\setminus v_k(f')}\cdot 
q^{n-1-\frac{|v_k(f)|+|v_k(f')|}{2}}\cdot 
(1-q)^{\frac{|v_k(f)|+|v_k(f')|}{2}}.
\ee 
Finally from (\ref{graphsup.neigh}) and (\ref{graphslow.neigh}) 
we have the desired 
expression:
\bd
(\om\cdot\overline{\om})_k(f,f')=q\cdot (\overline{\om}\cdot\om)_k(f,f')
\ed
\qed

\section{Fock States and Operations with Combinatorial Fock Spaces}
\label{Secoperations}
The operations between species of structures described in 
Section \ref{Secspecies} are helpful in understanding the 
action of creation and annihilation operators in terms of 
elementary ones. The guiding example is Green's representation of 
the operators appearing in parastatistics, as sums of bosonic 
(fermionic) operators with the ``wrong'' commutation relations
 \cite{Gre.}. Similar ideas appear in \cite{Spe.} where the author 
considers macroscopic fields as linear combinations of basic bosonic
 fields with various commutation relations.\\ 

 Thus, the first question we address in this section  is the following:
 given two combinatorial Fock spaces $(F,\om_F)$ and $(G,\om_G)$, is 
there a natural natural weight associated to the species 
$F+G$, $F\cdot G$, $F\times G$, $F\circ G$ ? The second question is
 related to the notion of positive definite functions on pair partitions.
 A general theory of such functions has been introduced in 
\cite{Boz.Sp.1} in connection with the so called generalized
 Brownian motion. 
\subsection{Fock States}\label{subsecFock}
We will start with the latter question by introducing the necessary
 definitions. \\

\definitie
Let $S$ be a finite  ordered set. We denote by  $\mathcal{P}_2(S)$ 
is the set of pair partitions of $S$, that is 
$\mathcal{V}\in\mathcal{P}_2(S)$ if $\mathcal{V}=\{V_1,..,V_r\}$ 
where each $V_i$ is an ordered set containing two elements 
$V_i=(k_i,l_i)$ with $k_i,l_i\in S$, $k_i<l_i$ and $\{V_1,..,V_r\}$ 
is a partition of $S$ ($V_i\bigcap V_j=\emptyset$ for $i\neq j$ 
and $\bigcup_{i=1}^r V_i=S$).  The set of all pair partitions is
\bd
\mathcal{P}_2(\infty)=\bigcup_{r=1}^\infty\mathcal{P}_2 (2r).
\ed     

Let $\K$ be a Hilbert space. We denote by $\mathcal{C}_\K$ the 
$^ *$-algebra obtained from the free algebra with generators 
$c(f)$ and $c^*(f)$, ($f\in\K$) divided by the relations:  
\bd
c^*(\lambda f_1+ \mu f_2)=\lambda c^*(f_1)+\mu c^*(f_2), ~
\lambda,\mu\in \C, ~f_1,f_2\in\K,
\ed 
and 
\bd
c^*(f)=(c(f))^*.
\ed
We are interested in a particular type of positive functionals 
on $\mathcal{C}_\K$, called \textit{Fock states} 
\cite{Boz.Sp.1} which have the 
following expression on monomials of creation and annihilation operators:
\begin{equation}\label{fockstate}
\rho_t(c^{\sharp_1}(f_1)\cdot..\cdot c^{\sharp_n}(f_n))=
\left\{\begin{array}{ll}
 0 & \textrm{if $n$ odd} \\
\sum_{\mathcal{V}=\{V_1,..,V_{\frac{n}{2}}\}}\rho_t[V_1]\cdot..\cdot
\rho_t[V_{\frac{n}{2}}]\cdot t(\mathcal{V}) & \textrm{if $n$ even} 
\end{array}\right.
\end{equation}
\noindent
the sum running over all pair partitions $\mathcal{V}$ in 
$\mathcal{P}_2(2r)$, and the symbols $\sharp_i$ standing for creation 
or annihilation. For $V=(k,l)\in \mathcal{V}$
\bd
\rho_t[V]=\br f_k,f_l\ke\cdot Q(\sharp_k,\sharp_l)
\ed
with the 2 by 2 covariance matrix

\begin{displaymath}
Q=
\left(\begin{array}{ccc}
\rho(c_ic_i) & \rho(c_ic^*_i) \\
\rho(c^*_ic_i)  &  \rho(c^*_ic^*_i)
\end{array}\right)
= 
\left(\begin{array}{ccc}
 0 & 1 \\
 0 & 0
\end{array}\right).
\end{displaymath}
where $c_i=c(e_i)$ and $e_i$ is an arbitrary 
normalized vector in $\K$. \\
Let us consider a real subspace $\K_{\R}$ of $\K$ 
such that $\K=\K_{\R}\dsum i\K_{\R}$. 
The sub-algebra of $\mathcal{C}_\K$ generated by the ``field operators'' 
$\om (f)=c(f)+c^*(f)$ with $f\in\K_{\R}$ is denoted 
by $\mathcal{A}_\K$. If the restriction of the functional 
$\rho_t$ to the algebra 
$\mathcal{A}_\K\subset\mathcal{C}_\K$ is a state, then we call  
the function
\bd
t:\mathcal{P}_2(\infty)\ra \C.
\ed
\textit{positive definite} \cite{Boz.Sp.1}. 
In particular if $\rho_t$ is a state on $\mathcal{C}_\K$ then
 $t$ is positive definite. The converse is not true in general.
 We will show first that the vacuum 
state of a symmetric Hilbert space is an example of Fock state.

\begin{proposition}\label{prop:fock}
Let $(F, \om_F)$ be a combinatorial Fock space with 
$|F[\emptyset]|=1$. Let $\Omega_F$ be the vacuum vector
 of $\G_F(\K)$, $\K$ a Hilbert space. Then the functional 
$\rho_F(\cdot)=\br\Omega_F, \cdot\Omega_F\ke$ is a 
Fock state on $\mathcal{C}_\K$.
\end{proposition}
In order to prove this proposition we need to introduce one more tool.

\definitie
Let $\K$ be a Hilbert space, $(F, \om_F)$ a combinatorial Fock space 
and $A\in\mathcal{B}(\K)$ a bounded operator on $\K$. The 
\textit{second quantization} of $A$ is defined by
\begin{eqnarray}
d\G_F(A):\G_F(\K) &\ra &\G_F(\K)\nonumber \\
(d\G_F(A)\varphi)(g) & = & d\G(A)(\varphi (g))\nonumber
\end{eqnarray}
where the meaning of $d\G(A)$ on the right side is
\bd
d\G(A):\K^{\tens  n} \ra  \K^{\tens  n}
\ed
\bd
d\G(A):\psi_0\tens..\tens\psi_{n-1} \ra \sum_{k=0}^{n-1}
\psi_0\tens..\tens A\psi_k\tens..\tens\psi_{n-1}
\ed

\noindent\textbf{Remark.} The second quantization operator 
is a well defined operator on $\G_F(\K)$.  Indeed let 
$\varphi\in\G_F(\K)$ and $\tau\in\symm( n)$, then
\begin{eqnarray}
(d\G_F(A)\varphi)(F[\tau](g)) & = &  d\G(A)\varphi(F[\tau](g))\nonumber \\
 =d\G(A)\cdot U(\tau)\varphi (g) & =& U(\tau)(d\G_F(A)\varphi)(g)\nonumber
\end{eqnarray}
and thus $d\G_F(A)\varphi\in \G_F(\K)$. We have used the 
invariance of $d\G(A)$ under permutations:
\bd
d\G (A)U(\tau)= U(\tau)d\G (A). 
\ed 

\begin{lemma}
We have the following commutation relations:
\begin{eqnarray}
 ~[a(h), d\G_F(A)] & = & a(A^*h)  \label{comm.rel1}    \\ 
 ~[d\G_F(A), d\G_F (B)] & = & d\G_F([A,B])  \label{comm.rel2} 
\end{eqnarray}

\end{lemma}

\noindent\textit{Proof.} Let $\varphi\in\G_F(\K)$, 
$f\in F[ n],~g\in F[n+1]$. Then
\begin{eqnarray}
(a(h)d\G (A)\varphi)(f)  & = & 
\sum_{g}\om(f,g)\cdot 
\mathrm{inp}_n (h,(d\G(A)\varphi)(g))     \nonumber \\
& = & \sum_{g}\om(f,g)\cdot \mathrm{inp}_n (h,\sum_{k=0}^{n} 
\mathbf{1}\tens..\tens A\tens..\tens \mathbf{1}\varphi (g))\nonumber \\
& = & \sum_{g}\om(f,g)\cdot\sum_{k=0}^{n-1}
\mathbf{1}\tens..\tens A\tens..\tens\mathbf{1}~
\mathrm{inp}_n(h, \varphi (g))\nonumber \\
&&\quad  +\sum_{g}\om(f,g)\cdot\mathrm{inp}_n(h,
\mathbf{1}\tens..\tens A \varphi (g))\nonumber \\
& = & \left(\sum_{k=0}^{n-1}\mathbf{1}\tens..\tens A\tens..\tens 
\mathbf{1}\right)\sum_{g}\om(f,g)\cdot
\mathrm{inp}_n(h, \varphi (g))\nonumber  \\
&& \quad +\sum_{g}\om(f,g)\cdot
\mathrm{inp}_n(A^*h, \varphi (g)) \nonumber   \\
&=&d\G_F(A)a(h)+a(A^*h) \nonumber
\end{eqnarray}
which proves (\ref{comm.rel1}). The other commutator follows 
directly from the definition of the second quantization operator.

\qed 

\begin{lemma}\label{le.a.monomial} 
Let $\{e_j\}_{j\in J}$ be an orthonormal basis of $\K$ and 
denote $a_j^\sharp =a(e_j)^\sharp$. 
Then the following equation holds:
\be
a_i\prod_{k=1}^{n}a_{i_k}^{\sharp_k}\Omega = \sum_{k=1}^{n}
\delta_{i,i_k}\cdot\delta_{\sharp_k, *}\cdot a_{i_0}
\prod_{p=1}^{k-1}a_{i_p}^{\sharp_p}\cdot a_{i_0}^{*}\cdot
\prod_{q=k+1}^{n}a_{i_q}^{\sharp_q}\Omega   \label{a.monomial}
\ee
where the colors $(i_k)_{k=0,..,n}$ satisfy the property 
$i_k\neq i_0$ for all $k=1,..n$.
\end{lemma}
\noindent\textit{Proof.} For simplicity we denote 
$\Psi=\prod_{k=1}^{n}a_{i_k}^{\sharp_k}\Omega$. We notice that 
$a_{i_0}\Psi =0$ due to the assumption that $i_k\neq i_0$ for all 
$k=1,..n$. Then using (\ref{comm.rel1}) we get
\be
a_i\Psi=[a_{i_0}, d\G(|i_0 \rangle \langle i|)]\Psi =
a_{i_0} d\G(|i_0\rangle \langle i|)\Psi   \label{eq.dg}
\ee
By successively applying the following commutator  
\bd
[d\G(|i_0 \rangle \langle i|), a_{i_k}^{\sharp_k}]=
\delta_{i_k,i}\cdot\delta_{\sharp_k, *} \cdot a_{i_0}^{*} 
\ed
we obtain the sum in (\ref{a.monomial}).

\qed

\noindent\textit{Proof of Proposition \ref{prop:fock}.} It is clear 
that $\rho_F$ is a positive linear functional on $\mathcal{C}_\K$. 
We need to prove that 
it has the expression (\ref{fockstate}). From linearity of the creation 
operators and anti-linearity of the annihilation operators we conclude 
that it is sufficient to consider the vectors $f_i$ in (\ref{fockstate}) 
belonging to the chosen orthogonal basis. 
From
\bd
\rho_F(\prod_{k=1}^{n}a_{i_k}^{\sharp_k})=\br\Omega,
\prod_{k=1}^{n}a_{i_k}^{\sharp_k}\Omega\ke 
\ed
and considering the fact that the creation operator increases the level 
by one while the annihilation operator decreases it by one, we deduce 
that nonzero expectations can appear only if $n$ is even and the 
number of creators in the monomial $\prod_{k=1}^{n}a_{i_k}^{\sharp_k}$ 
is equal to that of annihilators. Furthermore $a_{i_1}^{\sharp_1}$ 
must be an annihilator and $a_{i_n}^{\sharp_n}$, a creator. We will 
thus consider that this is the case.\\
We put the monomial in the form $a_{i_1}\prod_{k=2}^{n}a_{i_k}^{\sharp_k}$
 and  apply lemma \ref{le.a.monomial}. We obtain a sum over all pairs 
$(a_{i_1}, a_{i_k}^*)$ of the same color ($i_1=i_k$) and replace $i_1$ by
 a new color $i_0$. We go now to 
the next annihilator in each term of the sum and repeat the procedure, 
the new color which we add this time being different from all the 
colors used previously. After $\frac{n}{2}$ steps we obtain a sum
 containing all possible pairings of annihilators and creators
 of the same color in $ \prod_{k=1}^{n}a_{i_k}^{\sharp_k}$:

\bd
\rho_F(\prod_{k=1}^{n}a_{i_k}^{\sharp_k})= 
\sum_{\mathcal{V}=\{V_1,..,V_\frac{n}{2}\}}~\prod_{p=1}^{\frac{n}{2}}
\delta_{i_{k_p},i_{l_p}}\cdot Q(\sharp_{k_p},
\sharp_{l_p})\cdot t(\mathcal{V}) 
\ed  

with $V_p=(k_p,l_p)$ and $t(\mathcal{V})$ is given by
\bd
t(\mathcal{V})=\rho_F(\prod_{k=1}^{n}a_{j_k}^{\sharp_k})
\ed
such that $j_{k_p}=j_{l_{p'}}$ if and only if $p=p'$, for 
$p, p'\in\{1,..,\frac{n}{2}\}$, $\sharp_{k_p}$ is annihilator 
and $\sharp_{l_p}$ is creator.

\qed

\noindent
Thus for each combinatorial Fock space $(F, \om_F)$
 (which has a vacuum), the vacuum state is described by a 
positive definite function $t_F$ on $\mathcal{P}_2(\infty)$.  \\

\noindent\textbf{Remark.} We observe that the result can be
 generalized to a larger range of states and monomials. Let us 
partition the index set $J$ of the orthonormal basis of
 the Hilbert space $\K$ 
\bd
J=J_1+J_2
\ed
and choose a state $\rho_{\Phi}(\cdot)=\br\Phi,\cdot\Phi\ke$ and 
monomials $\prod_{k=1}^{n}a_{j_k}^{\sharp_k}$ such that  
$j_k\in J_1$ and $\Phi\in \G_F(\K_2)\subset\G_F(\K)$ is a 
normalized vector where $\K_2$ is the subspace of $\K$ with 
the orthogonal basis $\{e_j\}_{j\in J_2}$. Then it is easy to 
see that the argument used in the above proof still holds and 
$\rho_{\Phi}$ is a Fock state for $\mathcal{C}_{\K_1}$. In general 
$\rho_{\Phi}$ and $\rho_F$ do not coincide. When they do coincide we say 
that $\rho_F$ has the \textit{pyramidal independence} 
property \cite{Boz.Sp.1}.   

\subsection{Operations with symmetric Hilbert Spaces}
We pass now to the first question which we have posed in the 
beginning of this section. The various operations between species
 offer the opportunity of creating new symmetric Hilbert spaces 
which sometimes give rise to interesting interpolations between 
the two members. For the definitions of the operations we refer back
 to Section \ref{Secspecies}.\\

\noindent\textbf{1) Sums.} Let $(F,\om_F)$ and $(G,\om_G)$ be two 
combinatorial Fock spaces. From Section \ref{Secanfunct} we know that
\bd
\G_{F+G}=\G_F\oplus \G_G.
\ed
Note that the vacuum of $\G_{F+G}$ has dimension $\geq 2$ if 
$F[\emptyset]\neq\emptyset\neq G[\emptyset]$. The natural 
weight on $F+G$ is 
\bd
\om_{F+G}(t_1,t_2)=\om_F(t_1,t_2)+\om_G(t_1,t_2)
\ed
which gives rise to operators
\bd
a_{F+G}(h)=a_F(h)\oplus 0+0 \oplus a_G (h).
\ed
We consider a linear combination of the two vacua 
(for $|F[\emptyset]|=|G[\emptyset]|=1)$
\bd
\Omega_{\lambda}=\sqrt{\lambda}\Omega_F+\sqrt{1-\lambda}\Omega_G.
\ed
 The corresponding state $\rho_{F+G,\lambda}(\cdot)=
\br\Omega_{\lambda},\cdot\Omega_{\lambda}\ke$ interpolates 
linearly between $\rho_{F}$ and $\rho_{G}$
\bd
\rho_{F+G,\lambda}=\lambda\rho_{F}+(1-\lambda)\rho_{G}
\ed
and the same is true for the positive definite functions
\be\label{eq.product}
t_{F+G,\lambda}=\lambda t_F+(1-\lambda)t_G.
\ee

\noindent\textbf{1) Products.} Let $(F,\om_F)$ and $(G,\om_G)$ 
be two combinatorial Fock spaces. We consider the product 
species $F\cdot G$. As we have proved in Section \ref{Secanfunct}, 
there is the following isomorphism
\be\label{eq.tensorproduct}
\G_{F\cdot G}(\K)=\G_{F}(\K)\tens\G_{G}(\K).
\ee
Again there is a natural weight for the species $F\cdot G$. 
For $f\in F[U_1], g\in G[U_2], f'\in F'[U_1], g'\in G'[U_2]$:
\begin{eqnarray}
\om_{F\cdot G,\lambda}((f, g), (f,g')) 
& = & \sqrt{\lambda}\om_G(g,g') \nonumber \\
\om_{F\cdot G,\lambda}((f, g), (f',g)) 
& = &  \sqrt{1-\lambda}\om_F(f,f')\nonumber
\end{eqnarray}
all other values of $\om_{F\cdot G,\lambda}$ being 0. \\
From (\ref{eq.tensorproduct}) and the expression of 
$\om_{F\cdot G}$ we obtain
\bd
a_{F\cdot G}^\sharp (h)=\sqrt{\lambda}a_F^\sharp (h)
\tens\mathbf{1}+\sqrt{1-\lambda}\mathbf{1}\tens a_G^\sharp (h)
\ed
If $|F[\emptyset]|=|G[\emptyset]|=1$ then the
 state $\rho_{F\cdot G}(\cdot)=
\br\Omega_F\tens\Omega_G,\cdot \Omega_F\tens\Omega_G\ke$ 
generates the positive definite function:
\bd
t_{F\cdot G}(\mathcal{V})=
\sum_{\mathcal{V}_1,\mathcal{V}_2} \lambda^{|\mathcal{V}_1|}
(1-\lambda)^{|\mathcal{V}_2|}t_{F}(\mathcal{V}_1)
\cdot t_{G}(\mathcal{V}_2)
\ed
where the sum runs over all partitions of $\mathcal{V}$ 
in two sets, $\mathcal{V}_1$ and $\mathcal{V}_2$.\\

\noindent\textbf{Example:} The Green representation \cite{Gre.} 
of the (Fermi) parastatistics of order p is an example of 
application of the product of species. We consider the p-th 
power $(E_{\pm})^p$ of the species of oriented sets $E_{\pm}$. 
Then the annihilation operators are 
\bd
a(h)=\frac{1}{\sqrt{p}}\sum_{k=1}^p a^{(k)}(h)
\ed
and the vacuum state  is $\rho(\cdot)=\br\Omega,\cdot \Omega\ke$
 where $a^{(k)}$ is the term corresponding to the k-th term in 
the product and
\bd
\Omega = \Omega^{(1)}_a\tens ..\tens \Omega^{(p)}_a
\ed
is the tensor product of the antisymmetric vacua of each of 
the species $E^{(k)}_{\pm}$.\\

\noindent\textbf{3) Cartesian Products.}  Let $(F,\om_F)$ and 
$(G,\om_G)$ be two combinatorial Fock spaces. We consider the
 cartesian product species $F\times G$. The corresponding 
weight has the expression:
\bd
\om_{F\times G}((f,g), (f',g'))=\om_F(f,f')\cdot\om_G(g,g')
\ed 
We note that $\om_{F\times G}$ satisfies the invariance condition
 stated in the definition of the weight. 

\begin{proposition} \label{Prop.cartprod}
Let $(F,\om_F)$ and $(G,\om_G)$ be two combinatorial Fock spaces
both having a single structure on $\emptyset $. Then the
 positive definite function associated to the vacuum state of 
$(F\times G, \om_{F\times G})$ satisfies:
\be\label{eq.cartprod}
t_{F\times G}(\mathcal{V})=t_F(\mathcal{V})\cdot t_G(\mathcal{V})
\ee 
for all $\mathcal{V}\in \mathcal{P}_2(\infty)$.
\end{proposition} 

\noindent\textit{Proof.} We construct a linear operator $T$ from 
$\G_{F\times G}(\K)$ to $\G_F(\K)\tens\G_G(\K)$ with the property 
that its restriction to a certain subspace $\G_{F\times G}^\mathrm{ext}$ 
of $\G_{F\times G}(\K)$, is an isometry. The subspace 
$\G_{F\times G}^\mathrm{ext}$ is spanned by vectors $\delta_{[(f,g),c]}$ 
of the orthogonal basis $(F\times G)(J)$ of $\G_{F\times G}(\K)$ 
which have all colors different from each other, i.e. 
$c(i)\neq c(j)$ for $i\neq j$. We refer to Section \ref{Secanfunct} 
for the definitions related to the orthogonal basis 
of $\G_{F\times G}(\K)$.\\ 
\noindent
The action of $T$ on the basis vectors is:
\begin{eqnarray}
T:\G_{F\times G}^\mathrm{ext}(\K) & \ra & 
\G_F^\mathrm{ext}(\K)\tens\G_G^\mathrm{ext}(\K)          \nonumber\\
\delta_{[(f,g),c]} & \mapsto & \delta_{[f,c]}\tens\delta_{[g,c]}\nonumber
\end{eqnarray}
We check that the operator is well defined. Indeed the map
\begin{eqnarray}
i:\sum_{n=0}^{\infty}(F\times G)[ n]\times J^{ n}   & \ra & 
(\sum_{n=0}^{\infty}F[ n]\times J^{ n})\times
(\sum_{n=0}^{\infty}G[ n]\times J^{ n})           \nonumber\\
((f,g),c) & \mapsto & ((f,c), (g,c))                  \nonumber
\end{eqnarray}
commutes with the action of $\symm ( n)$ on the two sides, 
at each level and thus projects to a well defined map on the quotient:
\begin{eqnarray}
~i:(F\times G)(J) & \ra & F(J)\times G(J)                    \nonumber\\
~[(f,g),c] & \mapsto & ([f,c], [g,c])                            \nonumber
\end{eqnarray}
This means that $T$ is well defined. But as we have shown in
 Section \ref{Secanfunct}, the vectors $\delta_{[(f,g), c]}$,
 $\delta_{[f, c]}$ and $\delta_{[g, c]}$ for which $c(i)\neq c(j)$ 
if $i\neq j$, have norm one which implies that $T$ is an isometry. \\
Let us now consider the vector
\bd
\varphi_F^{(p)} = \prod_{k=1}^{p} a_{F,i_k}^{\sharp_k}~\Omega_F
\ed
the colors $(i_k)_{k=1,..,n}$ satisfying the condition that there
 are no three identical colors, and if there exists $k_1<k_2$ such
 that $i_{k_1}=i_{k_2}$, then $a_{i_{k_1}}^{\sharp_{k_1}}=a_{i_{k_1}}$
 and $a_{i_{k_2}}^{\sharp_{k_2}}=a_{i_{k_2}}^*$. It is clear that
 $\varphi_F^{(p)} \in \G_F^\mathrm{ext}(\K)$. Analogously we define
 $\varphi_G^{(p)}$ and  $\varphi_{F\times G}^{(p)}$. We want to 
prove by induction w.r.t. $p$ that the action of the isometry $T$ 
is such that
\be\label{induction}
T:\varphi_{F\times G}^{(p)}\ra  \varphi_F^{(p)} \tens \varphi_G^{(p)}.
\ee
This implies in particular (\ref{eq.cartprod}), when the monomial 
$\prod_{k=1}^{p} a_{i_k}^{\sharp_k}$ contains equal number of creators 
and annihilators pairing each other according to color, no two pairs
 having the same color. \\
\noindent
For $p=0$ we have $T(\Omega_{F\times G})=\Omega_F\tens\Omega_G$. 
Suppose (\ref{induction}) holds for p. Then there are two possibilities
 for increasing the length of the monomial 
$\prod_{k=1}^{p} a_{i_k}^{\sharp_k}$ by one: either by adding on the
 first position a creation operator $a^*_{i_0}$ such that the color
 $i_0$ does not appear in the rest of the monomial, or by adding an
 annihilation operator $a_{i_0}$ such that the term $a_{i_0}^*$ appears
 once in the rest of the monomial. We treat the two cases separately.\\
\noindent 1.) suppose that we have $\varphi_{F\times G} ^{(p)} =
 \prod_{k=1}^{p} a_{i_k}^{\sharp_k}~\Omega_{F\times G}$, $i_0\neq i_k$ 
 which has the decomposition 
\bd
\varphi_{F\times G} ^{(p)} = \sum_{[(f,g),c]}
\varphi([(f,g),c])\delta_{[(f,g),c]} 
\ed
with $\varphi([(f,g),c])\in \C$. Then 
\bd
a_{F\times G, i_0}^*\varphi_{F\times G} ^{(p)} =
\sum_{[(f,g),c],(f'g')}\varphi([(f,g),c])\cdot
\om_{F\times G}((f,g), (f',g'))~\delta_{[(f',g'),c^+_{i_0}]}
\ed
which implies 
\begin{eqnarray}
T(a_{F \times G,i_0}^*\varphi_{F\times G}^{(p)}) 
& = &  \sum_{[(f,g),c],(f'g')}\varphi([(f,g),c])~a_{F, i_0}^*
\delta_{[f,c]}\tens a_{G, i_0}^*\delta_{[g,c]} \nonumber \\
 = a_{F, i_0}^*\tens a_{G, i_0}^* T(\varphi_{F\times G}^{(p)}) 
& = &  a_{F, i_0}^* \varphi_F^{(p)}\tens a_{G, i_0}^* 
\varphi_G^{(p)}.     \nonumber 
\end{eqnarray}
\noindent 2.) suppose that we have 
$\varphi_{F\times G}^{(p)} = \prod_{k=1}^{p} a_{i_k}^{\sharp_k}~
\Omega_{F\times G}$ such that the term $a_{i_0}^*$ appears 
exactly one time in the the monomial 
$\prod_{k=1}^{p} a_{i_k}^{\sharp_k}$. We use again 
the Fourier decomposition
\bd
\varphi_{F\times G} ^{(p)} = \sum_{[(f,g),c]}
\varphi([(f,g),c])\delta_{[(f,g),c]} 
\ed
and identify in each orbit $[(f,g),c]\in (F\times G)(J)$, 
a representant $((f,g),c)\in (F\times G)[ n]\times J^{ n}$ 
such that $c(n-1)=i_0$. Then 
\bd
a_{F\times G, i_0}\varphi_{F\times G}^{(p)} =\sum_{[(f,g),c],(f'g')}
\varphi ([(f,g),c])\cdot\om_{F\times G}((f',g'), (f,g))~
\delta_{[(f',g'),c^-_{i_0}]}
\ed
where $c^-_{i_0}$ is the restriction of $c$ to the set
 $n-1$. Finally
\begin{eqnarray}
 T(a_{F\times G, i_0}\varphi_{F\times G}^{(p)}) 
& = &  \sum_{[(f,g),c],(f'g')}\varphi ([(f,g),c])\cdot
\om_{F\times G}((f',g'), (f,g))~T(\delta_{[(f',g'),c^-_{i_0}]}) \nonumber \\
 &= & \sum_{[(f,g),c],(f'g')}\varphi ([(f,g),c])\cdot
\om_F(f,f')\cdot\om_G(g,g') \delta_{[f',c^-_{i_0}] }
\tens \delta_{[g',c^-_{i_0}] } \nonumber \\ 
 & = & a_{F, i_0}\varphi_{F}^{(p)}\tens a_{G, i_0}
\varphi_{G}^{(p)} \nonumber
\end{eqnarray}
which proves the induction hypothesis for $p+1$ and the proposition.

\qed  \\
\noindent\textbf{Application:} Combining the result of the previous
 proposition and certain variations on the species of rooted
 trees, we investigate more general commutation relations of the type:
\bd
[a(h_1), a^*(h_2)]=\br h_1, h_2\ke \cdot f(N)
\ed  
with $f:\mathbf{N}\ra \mathbf{R}$ and $N$ the number operator
 characterized by 
\bd
[N,a^*(h)]=a^*(h).
\ed  
\begin{theorem}\label{th.polycomm}
Let $P$ be a real polynomial with positive coefficients. Then
 the commutation relations
\bd
[a(h_1), a^*(h_2)]=\br h_1, h_2\ke \cdot P(N)
\ed
are realizable on a symmetric Hilbert space.
\end{theorem}

\noindent We split the proof in a few lemmas.
\begin{lemma}
Let $(F,\om_F)$ and $(G,\om_G)$ be two symmetric Hilbert spaces 
for which the commutation relations hold
\begin{eqnarray}
~[a_F(h_1), a^*_F(h_2)] & = & \br h_1, h_2\ke \cdot a(N)\nonumber \\
~[a_G(h_1), a^*_G(h_2)] & = & \br h_1, h_2\ke \cdot b(N)\nonumber
\end{eqnarray}
where $a,b$ are real functions. Then on 
$(F\times G, \om_{F\times G})$ we have 
\bd
[a_{F\times G}(h_1), a^*_{F\times G}(h_2)] 
=\br h_1, h_2\ke \cdot (a\cdot b)(N)
\ed
\end{lemma}
\noindent\textit{Proof.} This is a direct application 
of Lemmas \ref{a*a}, \ref{aa*} and the following equations:
\begin{eqnarray}
(\om\cdot\overline{\om})_k((f,g),(f',g')) 
& = & (\om\cdot\overline{\om})_k(f,f')\cdot 
(\om\cdot\overline{\om})_k(g,g')\nonumber \\
(\overline{\om}\cdot\om)_k((f,g),(f',g')) 
& = & (\overline{\om}\cdot\om)_k(f,f')\cdot 
(\overline{\om}\cdot\om)_k(g,g')\nonumber
\end{eqnarray}

\qed

\begin{lemma} Let $\mathcal{A}$ be the species of rooted trees. 
Let $f\in\mathcal{A}[U]$, \\
$g\in\mathcal{A}[U+\{*\}]$ and 
\bd
\tilde{\om}_{\mathcal{A}}^c(f,g)=\om_\mathcal{A}(f,g) 
+ c^\frac{1}{2}\delta_{f_*,g}
\ed
a modification of the weight $\om_\mathcal{A}$ defined 
in section \ref{Secalgebras}, with $c$, a positive constant.
 The structure $f_*\in \A[U+\{*\}]$ is defined by: 
\begin{displaymath}
f_*(u)=\left\{\begin{array}{ll}
f(u) & \textrm{if $u\neq \textrm{root}(f)$}\\
\mathrm{*}  & \textrm{if $u = \textrm{root}(f)$.}
\end{array}\right.
\end{displaymath}
Then on $(\A, \tilde{\om}_\A^c)$ we have:
\bd
[a(h_1), a^*(h_2)]=\br h_1, h_2\ke \cdot (N+c).
\ed
\end{lemma}
\noindent\textit{Proof.} This is similar to the proof of 
Theorem \ref{th.rootedtrees}, with an additional contribution 
to $(\om\cdot\overline{\om})_n(f,g)$ coming from the term 
$ c^\frac{1}{2}\delta_{f_*,g}$ 
in $\tilde{\om}_{\mathcal{A}}^c$. 

\qed

\begin{lemma} Let $\mathcal{A}\times\mathcal{A}$ be the 
species of ordered pairs of rooted trees. We define the weight
\bd
\om_{\A\times\A}^c((f,g),(f',g'))=\om_\A(f,g)\cdot
\om_\A(f',g')+ c^\frac{1}{2}\delta_{f_*,f'}\cdot\delta_{g_*,g'}.
\ed
Then on  $(\A\times \A, \om_{\A\times\A}^c)$ we have
\bd
[a(h_1), a^*(h_2)]=\br h_1, h_2\ke \cdot (N^2+c).
\ed
\end{lemma}

\noindent\textit{Proof.} Similar to the previous two lemmas.

\qed

\noindent{Proof of Theorem \ref{th.polycomm}.} The polynomial 
$P(x)$ has a canonical expression as product of polynomials 
of the type $x+c$ and $x^2+c$ with $c\geq 0$. The theorem 
follows by applying the previous 3 lemmas.

\qed

\noindent\textbf{Remark.} The result can be extended to power
 series with positive coefficients and infinite radius of convergence. 
In particular for $ 0\leq q \leq 1$
\bd
s(x)=q^{-x}=\sum_{k=0}^\infty \frac{1}{n!}\cdot (-\log q)^n\cdot x^n
\ed  
gives the commutation relations
\bd 
[a_i, a^*_j]=q^{-N}\delta_{i,j}
\ed
which characterize the \textit{$q$-deformations} 
\cite{Boz.Sp.2}, \cite{Grb.},
 up to a ``rescaling'' of the creation and annihilation operators 
with a function of $N$.\\

\noindent\textbf{4) Compositions.} Let $(F,\om_F)$ and $(G,\om_G)$ 
be two combinatorial Fock spaces. We recall that the composition 
of $G$ in $F$ is a species whose structures 
are $F$-assemblies of $G$-structures:
\bd
F\circ G[U]=\sum_{\pi} F[\pi]\times \prod _{p\in \pi} G[p].
\ed 

 We would like to define the annihilation and creation operators for 
the species $F\circ G$ by making use of the available weights $\om_F$ 
and $\om_G$. Apart from the condition $|G[\emptyset]|=0$ we require 
$|G[1]|=1$. We consider an arbitrary structure 
$(f,\pi, (g_p)_{p\in\pi})\in F\circ G[U]$ where $\pi$ is a partition
 of the finite set $U$. Then we note that there are two essentially
 different possibilities to ``add'' a new point $*$, to the set $U$: one
 can enlarge the size of $\pi$ by creating a partition of $U+\{*\}$ of
 the form $\pi^+=\pi +\{\{*\}\}$, or one can keep the size of $\pi$ constant
 by adding $*$ to one of the sets $p\in\pi$ and obtain the 
partition $\pi^+_p$. Between $\pi$ and $\pi^+_p$ there is the bijection
\bd
\alpha_p: p'\ra \left\{ \begin{array}{ll}
p' & \textrm{if $p'\neq p $}\\ 
p+\{*\} & \textrm{if $p'=p $}
\end{array}\right.
\ed

\noindent We recognize that in the first case the weight $\om_F$
 should play a role, while in the second, the weight $\om_G$. According
 to the properties of the species $F$, one can further distinguish
 among the subsets to which $*$ is added, by choosing (as we did for
 the creation and annihilation operators) a weight $\om_{F,\epsilon}$
 on the cartesian product $F\times \epsilon$ where $\epsilon$ is the
 species of elements: $\epsilon[U]=U$. Putting together the three
 data $(\om_F, \om_G, \om_{F, \epsilon})$, we define:
\begin{eqnarray}
&  & \om_{F\circ G}((f, \pi, (g_p)_{p\in \pi}),
 (f', \pi', (g'_{p'})_{p'\in \pi'})) = \om_F(f,f')\cdot
\prod_{p\in\pi}\delta_{g_p, g'_p}  \nonumber \\
& + & \sum_{p\in \pi} \delta_{f', F[\alpha_p](f)}\cdot 
\om_{F, \epsilon}(f,p)\cdot \om_G(g_p, g'_{p+\{*\}}).\nonumber
\end{eqnarray}
where $f\in F[\pi],~ g_p\in G[p]$, etc.

\noindent\textbf{Remark.}We find this definition rather natural and
 broad enough to cover some interesting examples. One can easily
 check that $\om_{F\circ G}$ satisfies the invariance property
 characterizing the weights.\\
\noindent\textbf{Example:} The species $\textrm{Bal}$ of ordered
 partitions or Ballots is the composition of
 $L$ (the species of linear orderings ), with $E_+$ 
(the species of nonempty sets). A typical structure over a
 finite set $U$ looks like: $s=(U_1,..,U_k)$ with 
$(U_p)_{p\in\{1,..,k\}}$, a partition of $U$. The vacuum is the empty
 sequence $\textrm{Bal}[\emptyset]=\{\emptyset\}$. We use the
 weights $\om_E$ and $\om_L$ as defined in section \ref{Secalgebras}.
 The action of the creation operator at the combinatorial
 level can be described as follows: we can add the point $*$ in the
 last subset of the sequence $s=(U_1,..,U_k)$ and obtain 
$s^+_k=(U_1,..,U_k+\{*\})$, or we can create a new
 subset $U_{k+1}=\{*\}$ and position 
it at the end of the sequence $s$, producing $s^+=(U_1,..,U_{k+1})$.
 We see that in this case the weight $\om_{L,\epsilon}$ is simply 
identifying the last element of the sequence:
$\om_{L,\epsilon}(s,U_k)=\delta_{U_k,U_p}$. For the vacuum we
 set $\om_ \mathrm{Bal}(\{\emptyset\},\{*\})=1$. We use
 $0\leq q\leq 1$ as an interpolation parameter:
\be\label{eq.omegabal}
\om_\mathrm{Bal}(s,s')=q^\frac{1}{2}
\delta_{s^+_k,s'}~+~(1-q)^\frac{1}{2}\delta_{s^+,s'} 
\ee
Let us denote by $t_\mathrm{Bal}$ the positive definite
 function associated to the vacuum state of the 
combinatorial Fock space $(\textrm{Bal}, \om_\mathrm{Bal})$,
 as defined in subsection \ref{subsecFock}. 
Following \cite{Boz.Sp.1} we associate to any pair partition
 $\mathcal{V}\in\mathcal{P}_2(\infty)$ a set  
$B(\mathcal{V})=\{\mathcal{V}_1,..,\mathcal{V}_k\}$ such that 
$\mathcal{V}=\mathcal{V}_1\cup...\cup\mathcal{V}_k$ is 
the decomposition of $\mathcal{V}$ into connected 
sub-partitions or \textit{blocks}.
\begin{theorem}
Let $\mathcal{V}\in\mathcal{P}_2(\infty)$. Then
\be\label{eqbal}
t_\mathrm{Bal}(\mathcal{V})=q^{|\mathcal{V}|-|B(\mathcal{V})|}
\ee
\end{theorem}
\noindent\textit{Proof.} We split the task of proving 
(\ref{eqbal}) into two simpler ones: first we prove the 
\textit{strong multiplicativity} property for $t$:
\bd
t(\mathcal{V})=\prod_{i=1}^{k} t(\mathcal{V}_i)~~ \mathrm{if}~~ 
B(\mathcal{V})=\{\mathcal{V}_1,..,\mathcal{V}_k\}
\ed
 and then for $\mathcal{V}$ consisting of a single 
block, $t_\mathrm{Bal}(\mathcal{V})=q^{|\mathcal{V}|-1}$. 
The proof of the strong multiplicativity is analogous to that of 
Proposition \ref{Prop.cartprod}. We consider an orthogonal
 basis $(e_j)_{j\in J}$ of the Hilbert space $\K$ and a partition
 $J=J_1+J_2$ of $J$ with the corresponding relation 
$\K=\K_1\oplus\K_2$. We define an isometry
\bd
S:\G_\mathrm{Bal}^\mathrm{ext}(\K_1)\tens\G_\mathrm{Bal}^\mathrm{ext}(\K_2) 
 \ra  \G_\mathrm{Bal}^\mathrm{ext}(\K_1\oplus\K_2)
\ed
 and we will prove that it has a natural action on monomials 
of creation and annihilation operators:
\be\label{eq.prmon}
S(\prod_{k}a^{\sharp_k}_{i_k}\Omega \tens 
\prod_{p}a^{\sharp_p}_{j_p}\Omega )=
\prod_{k}a^{\sharp_k}_{i_k} \cdot \prod_{p}a^{\sharp_p}_{j_p}\Omega.
\ee

\noindent We recall that the two monomials satisfy certain 
properties which are described in Proposition \ref{Prop.cartprod}.
 The multiplicativity of $t_\mathrm{Bal}$ follows from 
equation (\ref{eq.prmon}) and the isometric property of $S$.\\
\noindent
The action of $S$ on the orthogonal bases defined in Section 
\ref{Secanfunct} is:
\bd             
\delta_{[s_1,c_1]} \tens \delta_{[s_2,c_2]} 
 \ra  \sum_{s} q^\frac{a(s)}{2} \cdot (1-q)^\frac{b(s)}{2} \cdot
 \delta_{[s,c]}\ed

\noindent where, for arbitrary $s_1=(U_1,..,U_k)$ and 
$s_2=(V_1,..,V_p)$, the sum runs over all 
$s=(V_1,..,V_{p-1},~V,~ U,~ U_2,..,U_k)$ with 
$V_p\subset V,~U\subset U_1$ and $U+V=U_1+V_p$. The coloring $c$ restricts 
to $c_1$ and $c_2$ on the sets $\bigcup_{\alpha} U_\alpha$ 
respectively $\bigcup_{\beta}V_\beta$. The coefficients appearing 
on the right side are $a(s)=|V|-|V_p|$ and $b(s)=|U|$. As 
$\|\delta_{[s,c]}\|=1$  and $a(s)+b(s)=|U_1|$, we obtain 
\begin{eqnarray}
\|S(\delta_{[s_1,c_1]} \tens \delta_{[s_2,c_2]})\|^2 & = & 
\sum_{j=0}^{|U_1|} {|U_1| \choose j}\cdot q^{k}\cdot 
(1-q)^{|U_1|-k}\nonumber \\
=1 & = & 
\|\delta_{[s_1,c_1]} \tens \delta_{[s_2,c_2]}\|^2, \nonumber 
\end{eqnarray}  

\noindent which proves the isometry property. 
The equation (\ref{eq.prmon}) follows by induction w.r.t. $k$.
 For $k=0$ is is obvious that
\bd
S\left(\Omega\tens \prod_{p}a^{\sharp_p}_{j_p}\Omega\right)=
\prod_{p}a^{\sharp_p}_{j_p}\Omega.
\ed
Then one can check on the basis vectors that 
\bd
S\cdot (a_j^{\sharp_j}\tens\mathbf{1})=a_j^{\sharp_j}\cdot S 
\ed
\noindent for $j\in J_1$, which provides the tool for the 
incrementation of $k$.

We pass now to the expression of $t_\mathrm{Bal}(\mathcal{V})$ 
for a one block partition $\mathcal{V}$. The basic observation 
is that the creation and annihilation operators have the following
 form, stemming from  that of $\om_\mathrm{Bal}$ 
(see (\ref{eq.omegabal}) ):
\bd
a_i^{\sharp_i}=q^\frac{1}{2}a_{E,i}^{\sharp_i}+(1-q)^
\frac{1}{2}a_{L,i}^{\sharp_i}
\ed
with the choice $a_{E,i}^*\Omega=0$. Let 
$M_\mathcal{V}=\prod_{l=1}^{2n}a_{i_l}^{\sharp_l}$ 
be a monomial associated to the pair partition 
$\mathcal{V}\in\mathcal{P}_2(2n)$. It is sufficient to prove that
 the only nonzero contribution to $M_\mathcal{V}\Omega$ is brought
 by the the term $a_{L,i_1} \prod_{l=2}^{2n-1}a_{E,i_l}\cdot 
a^*_{L,i_{2n}}\Omega=q^{n-1}\Omega$. Indeed the action of 
$a^*_{L,i_1}$ at the combinatorial level is to increase the number
 of subsets in a sequence by 1. Thus the terms which are nonzero
 must contain an equal number of creation and annihilation
 operators of type $L$. Let us consider such a term. Then there
 exist $1\leq l_1\leq l_2\leq 2n$ such that on the positions $l_1$
 and $l_2$ we have annihilation respectively creation operators
 of type $L$ and for $l_1\leq l\leq l_2$ we have type $E$ operators.
 We have identified a submonomial 

\bd
m=a_{L, l_1}\cdot\prod_{l=l_1+1}^{l_2-1}a_{E,i_l}^{\sharp_l}\cdot 
a_{L,l_2}^*
\ed
which is nonzero only if it corresponds to a pair partition,
 that is if all creation and annihilation operators pair each 
other according to the color. But this is possible only when 
$l_1=1$ and $l_2=2n$ because  $\mathcal{V}$ is a one-block pair partition.\\

\qed

\noindent\textbf{4) Free Products.} Inspired by the notion of
 freeness introduced by \\
 Dan Voiculescu \cite{Voi.Dy.Ni.} we make the following:\\
\definitie
Let $(F_\alpha)_{\alpha\in J}$ be a finite set of species of
 structures with \\$F_\alpha[\emptyset]=\{\emptyset\}$ for all
 $\alpha\in J$. The \textit{free product} of $(F_\alpha)_{\alpha\in J}$ 
is the species defined by:
\begin{eqnarray}
 *_{\alpha\in J}(F_\alpha)[U] & = &
 \{(\pi, (s_1,..,s_p))|\pi=(U_1,..,U_p)\in \mathrm{Bal}[U], 
s_i\in F_{\alpha_i}[U_i],\nonumber \\
& &  \alpha_i\neq\alpha_{i+1} ~\textrm{for}~i=1,..p-1\} 
\end{eqnarray}
for $U\neq\emptyset$ and $*_{\alpha\in J}F_\alpha[\emptyset]=
\{\emptyset\}$. The transport is induced from the species 
$(F_\alpha)_{\alpha\in J}$ and $\textrm{Bal}$. 
From the definition it is clear that we have the following 
combinatorial equation:
\bd
*_{\alpha\in J}(F_\alpha)=1+\sum_{p\geq 1}~
\sum_{\alpha_1\neq\alpha_2\neq..\neq\alpha_p} F_{\alpha_1+}\cdot 
F_{\alpha_2+} \cdot...\cdot F_{\alpha_p+}
\ed
and using the property $\G_{F\cdot G}=\G_F\tens\G_G$ we obtain
\bd
\G_{*_{\alpha\in J}(F_\alpha)}(\K)=*_{\alpha\in J}(\G_{F_\alpha},
 \Omega_\alpha)
\ed
where the last object is the Hilbert space free product \cite{Voi.Dy.Ni.}.\\
\noindent The corresponding weight is similar to the one
 used for the species Bal. For $f_i\in F_{\alpha_i}[U_i]$ 
and $f_i'\in F_{\alpha_i'}[V_i]$, it has the expression:
\begin{eqnarray}
& & \om_{*_{\alpha\in J(F_{\alpha})}}((\pi,(f_1,..,f_p)),
(\pi ',(f_1',..,f_q')))=\nonumber \\
&  & \delta_{p,q}\cdot\delta_{\alpha_p,\alpha_p'} 
\prod_{i=1}^{p-1}\delta_{f_i,f_i'}\cdot \om_{\alpha_p}(f_p.f_p')
+ \delta_{p+1, q}\cdot \prod_{i=1}^{p}\delta_{f_i,f_i'}\cdot 
\om_{\alpha_q'}(\{\emptyset\},f_q' ).\nonumber
\end{eqnarray}
Moreover the creation and annihilation operators can be written like
\bd
a^\sharp_{*_{\alpha\in J(F_{\alpha})},i}=
\sum_{\alpha}a^{\sharp}_{F_\alpha,i}
\ed
with the relations \cite{Voi.Dy.Ni.}, 
\bd
a_{F_\alpha,i}\cdot a^*_{F_\beta,j}=0
\ed
for $\alpha\neq\beta$.

\end{document}